\documentclass[fleqn,usenatbib]{mnras}

\usepackage{newtxtext,newtxmath}

\usepackage[T1]{fontenc}

\DeclareRobustCommand{\VAN}[3]{#2}
\let\VANthebibliography\thebibliography
\def\thebibliography{\DeclareRobustCommand{\VAN}[3]{##3}\VANthebibliography}

\usepackage{graphicx}	
\usepackage{amsmath}	
\usepackage{upgreek}
\usepackage{lscape}
\usepackage{array}
\usepackage{graphics}
\usepackage{tikz}
\usepackage{xcolor}
\usepackage{orcidlink} 
\usepackage{upgreek}
\usepackage{textcomp}
\usepackage{subfigure}
\usepackage{subcaption}
\usepackage{threeparttable}
\usepackage{makecell}

\newcommand{\arcs}{\ensuremath{^{\prime\prime}}}
\newcommand{\oldhtr}{CRAFT HTR1}
\newcommand{\newhtr}{CRAFT HTR2}
\newcommand{\rmeg}{|RM$_{\mathrm{EG}}$|}

\title[\newhtr]{\newhtr: Polarimetry of 64 non-repeating fast radio bursts from the updated CRAFT catalogue}

\author[T. Dial et al]{
T. Dial,$^{1}$\thanks{E-mail: tdial@swin.edu.au}
A. T. Deller, $^{1}$
Marcin Glowacki, $^{2,3,4}$
A. Bera, $^{3,5}$
R.~M.~Shannon, $^{1}$
\newauthor
Clancy W. James, $^{3}$
Joscha N.\ Jahns-Schindler \orcidlink{0000-0003-4193-6158}, $^{1}$
Ziteng Wang, $^{3}$
Akhil Jaini, $^{1}$
\\
$^{1}$Center for Astrophysics and Supercomputing, Swinburne University of Technology, P.O. Box 218, Hawthorn, Vic 3122 Australia\\
$^{2}$Institute for Astronomy, University of Edinburgh, Royal Observatory, Edinburgh, EH9 3HJ, United Kingdom\\
$^{3}$International Centre for Radio Astronomy Research, Curtin University, Bentley, WA 6102, Australia\\
$^{4}$Inter-University Institute for Data Intensive Astronomy, Department of Astronomy, University of Cape Town, Cape Town, South Africa\\
$^{5}$ASTRON, Netherlands Institute for Radio Astronomy, Postbus 2, 7990 AA Dwingeloo, Netherlands\\
}

\date{Accepted XXX. Received YYY; in original form ZZZ}

\pubyear{\the\year{}}

\begin{document}
\label{firstpage}
\pagerange{\pageref{firstpage}--\pageref{lastpage}}
\maketitle

\begin{abstract}

We present high-time resolution spectro-polarimetric data for 34 new fast radio bursts (FRBs) discovered by the  Commensal Real-time Fast Transients (CRAFT) survey on the Australian Square Kilometer Array Pathfinder (ASKAP) during the period May 2024 to June 2026. Most of these were detected by the higher-sensitivity CRAFT COherent (CRACO) detection system that was commissioned on the telescope during this period. This new sample doubles the size of the CRAFT HTR catalogue and probes a fainter population of FRBs thanks to the improved sensitivity of CRACO. We compare the distribution of extragalactic rotation measure (RM) and polarisation fraction to the CHIME and DSA catalogues. While no significant differences were seen between CRAFT and DSA, the extragalactic RM distribution seen in CHIME FRBs (which are detected at lower frequency) was substantially lower. Surprisingly, we find no significant differences in the linear polarisation fraction distribution between the three FRB catalogues, suggesting an indifference to the different telescope observing frequencies. We show tentative evidence for wider and fainter bursts possessing lower polarisation fractions; this is predominantly driven by the growing sample of unpolarised bursts that are, in almost all cases, wider ($\gg$10 ms) and fainter ($\ll$10$^{34}$ ergs~s$^{-1}$~Hz$^{-1}$) than the median ASKAP detection. 

\end{abstract}




\section{Introduction}

Fast Radio Bursts (FRBs) are extra-galactic millisecond bursts of radio emission \citep{lorimer2007bright}. FRBs are extremely bright, with spectral luminosities up to ten orders of magnitude greater than radio emission from Galactic pulsars \citep{petroff2022fast}, and are visible to high redshift \citep{caleb2025}. In the last decade they have proven great potential in probing free-electrons in the intergalactic medium (IGM), which help us in understanding the evolution of baryons and large-scale structure formation \citep{macquart2020census, baptista2024measuring, sharma2026}. However, because their observables provide information of integrated line-of-sight quantities, unlocking their potential to probe large scale structure is dependent on modelling their progenitors and surrounding environments. Furthermore, current observations suggest the possibility of two distinct populations of FRBs, repeating and non-repeating FRBs, which implies multiple progenitors, emission mechanisms, and/or environments. This severely impacts their potential as cosmological probes \citep{Wang_2026}.

 The high luminosities of FRBs may originate in violent emission mechanisms such as star quakes from neutron stars (NS) \citep{wu2025universal}. Bursts from repeating sources are typically one to three orders of magnitude less luminous than non-repeating FRBs \citep{kirsten2024link}. However, the accumulated energy release of the most active repeaters such as FRB 20201124A \citep{zhang2022fast} and FRB 20220912A \citep{zhang2023fast} start to breach the theoretical energy budgets of most compact objects such as magnetars \citep{zhang2025magnetar}, providing constraints on the assumptions of radio to bolometric luminosity in the progenitor.

Polarisation provides perhaps the most direct probe of the FRB emission mechanism, and hence the FRB progenitor population. FRBs typically have high levels of linear polarisation (L/I) and modest circular polarisation (V/I) \citep{sherman2024deep, pandhi2024polarization, scott2025high}. This may be intrinsic to the FRB population and potential evidence for emission mechanisms such as synchrotron radiation \citep{melrose1971degree}. On the other hand, instances of high levels of V/I have been seen in both repeating and non-repeating FRBs \citep{kumar2022, dial2025frb, jiang2025ninety} that could be either induced by propagation through the circum-burst environment \citep{beniamini2022faraday} or intrinsic to the emission mechanism \citep{zhang2023fast}. 

However, polarisation is subject to many propagation effects, including some that can be traced back to an extreme circum-burst environment. For instance, the rotation measure (RM) of the repeating FRB 20121102A has steadily declined by $\sim$15 per cent per year  \citep[after starting at the large magnitude of $\sim$10$^{5}$ rad m$^{-2}$,][]{michilli2018extreme} which could support the presence of a supernovae remnant \citep{hilmarsson2021rotation}. FRB 20190520B has shown RM sign reversal in its temporal evolution \citep{anna2023magnetic} and FRB 20220629A an abrupt spike in |RM| \citep{li2026sudden}, both of which may result from a binary companion. A handful of repeating sources also show evidence for spectral depolarisation \citep[e.g. FRB 20190520B and FRB 20201124A;][]{niu2022repeating, bruni2024nebular}. The environments of these repeating sources are often highly variable and magneto-ionised \citep{feng2022frequency}. Along with the presence of persistent radio sources (PRSs), there is strong evidence for young magnetars as the progenitors of these repeating sources \citep{bruni2025discovery}. 

The majority of bursts detected thus far come from a small number of the most active repeating sources -- which make up $<$1\% of the $\sim$4000 reported FRB sources to date \citep{chime2catalog} -- and have enabled us to place constraints on their individual progenitors and environments. They are often associated with PRSs \citep{niu2022repeating, Yang_2024} and have high RMs or found near complex magneto-ionised environments \citep{michilli2018extreme, anna2023magnetic, li2026sudden}. However, extrapolating these extreme examples to describe the broader FRB population may not be reliable. Therefore, recent studies have focused on the growing sample of non-repeating bursts with high-time resolution and full polarimetry. This includes 35 bursts from the commensal real-time fast transients \cite[CRAFT;][]{shannon2024commensal, scott2025high} survey which were detected using the Australian SKA pathfinder \citep[ASKAP;][]{hotan2021australian}, 118 bursts detected using Canadian Hydrogen Intensity Mapping Experiment \citep[CHIME;][]{CHIMEdesc} as part of the first CHIME/FRB catalogue \citep{amiri2021first, pandhi2024polarization} and 25 bursts detected with the Deep Synoptic Array \citep[DSA;][]{sherman2024deep}. In addition to the aforementioned polarisation properties, these samples show a handful of different temporal polarisation position angle (PA) archetypes, with most being either flat or exhibiting slow variations across the burst. The high time resolution afforded by ASKAP -- which can reach $\sim$3 ns when sacrificing frequency resolution \citep{james2025esd} but is typically $\sim$1 \textmu s when analysing the spectro-temporal properties of FRBs -- has enabled the study of bursts that display extremely narrow PA structures on \textmu s to sub-\textmu s timescales. Analysis of scattering times in the ASKAP burst sample suggest it is a dominant effect in masking these narrow PA structures that are intrinsic to the source and emission mechanism \citep{scott2025high}. At this point, the results published to date provide good evidence that at least some, if not all, non-repeating bursts originate from compact objects (e.g. magnetars) and propagate through circum-burst environments that can mask short timescale features and convert linear polarisation to circular polarisation. 

In this paper, we double the ASKAP burst sample to include an additional 34 bursts detected from May 2024 to June 2026. In Section \ref{sec:methods} we describe the detection and processing of these new bursts using the recently commissioned CRAFT COherent detection system \citep[CRACO;][]{wang2025craft} and improved CELEBI pipeline \citep{glowacki2026pinkupdateimprovementscelebi}. In Section \ref{sec:discussion} we discuss the polarisation features of the full ASKAP sample and compare these properties to those displayed by the DSA and CHIME samples. We analyse these polarisation features along with the burst energetics in the rest frame and find that brighter and narrower bursts are, on average, more linearly polarised than fainter and broader bursts. We note a growing sample of fainter unpolarised bursts that fit this correlation, and consider whether they can be explained by depolarisation in the circumburst medium. Finally we conclude our study in Section \ref{sec:conclusion}. 


\section{Methods}
\label{sec:methods}

Our processing proceeded identically to that detailed in the \oldhtr\ sample \citep{scott2025high}, with exceptions detailed below.

\subsection{FRB detection}
The new CRAFT COherent \citep[CRACO;][]{wang2025craft} detection system has been fully operational on ASKAP since May 2024, with 43 bursts being detected between this date and June 2026. Briefly, CRACO searches for short transient events by performing real-time millisecond imaging of the ASKAP field of view (FoV). For the CRACO system, visibilities from the ASKAP correlator are integrated to a 1 MHz frequency resolution (across a 288 MHz bandwidth) and a configurable time resolution before being de-dispersed, imaged, and searched. CRACO also searches integer multiples of the base time resolution to maintain sensitivity to longer bursts. For the bursts detected in this sample, CRACO was operating with time resolutions between 3.8 ms and 27.6 ms, with most bursts detected during operations with 13.8 ms resolution. Radio frequency interference (RFI) mitigation and candidate classification are performed on these images as is described in \cite{wang2025craft}. 

CRACO operates alongside the incoherent summation (ICS) FRB detection system \citep{shannon2024commensal}. While the CRACO system has a nominally higher sensitivity, it operates at a lower time resolution then the ICS system and with a somewhat smaller effective field of view (FOV). Out of the 43 bursts, 39 were detected by CRACO, of which 18 were also detected by the ICS system (generally at lower S/N). The remaining four bursts 20240525A, 20250613A, 20251024B and 20251026B were detected only by the ICS system, due to being either outside the CRACO FOV or being temporally very narrow (and hence lower S/N in CRACO than the ICS system). FRB 20250613A and 20251026B are bursts from the same repeating source FRB 20250613A and have already been described in detail in \cite{dial2026}.

The detection of an FRB triggers a dump of the 84 GB ring-buffer holding 12.4 s of 1+1 bit complex baseband voltage data per antenna. After the detection of an FRB, observations of a band-pass calibrator and polarisation calibrator are scheduled, during which 12.4 s of voltage data is also downloaded. 
The voltage data from the FRB and calibrators are transferred to the Ngarrgu Tindebeek supercomputer at
Swinburne University of Technology.

We employ one of two primary band-pass calibrators for each FRB, PKS B0407$-$658 and PKS B1934$-$638, depending on the local sidereal time. We also choose between  two polarisation calibrators depending on local sidereal time, PSR J0835$-$4510 (Vela) and PSR J1644$-$4559 (1644). Vela is the preferred polarisation calibrator, as discussed below. Neither source was available during calibrator observations following the detection of FRB 20251119B and FRB 20260614A. For the former, no polarisation calibration observation was taken, while for the later, PSR J0034$-$0721 was observed. However, as a broad-band polarisation model of this pulsar has not been generated from independent observations at this stage, neither of these two bursts have polarisation calibration solutions (described below) generated or applied.

Nine of the 43 bursts detected are not included in the remainder of this manuscript because of a lack of baseband data; this is due to either a failure in the voltage download, or an excessive latency in the trigger meaning the FRB fell outside the download window. This leaves 34 FRBs to add to the latest CRAFT sample. FRBs 20260217D and 20260307D had incomplete voltage downloads leaving only one polarisation beam. These bursts are still included in this sample for completeness and are shown in Fig.\ref{fig:dynspec_mosaic} but they are not analysed in Section \ref{sec:discussion}.

\subsection{Processing}

After the voltages are transferred to the Ngarrgu Tindebeek supercomputer, we run the CRAFT effortless localisation and enhanced burst inspection \cite[CELEBI;][]{scott2023celebi, glowacki2026pinkupdateimprovementscelebi} pipeline. The outputs of this pipeline are Stokes \textit{I} images of the FRB and the surrounding field that are used to determine the FRB position to high accuracy, along with full Stokes dynamic spectra produced by coherently dedispersing and beamforming the ASKAP data at that FRB position, as detailed in \cite{scott2023celebi} and \cite{glowacki2026pinkupdateimprovementscelebi}. The default frequency and time resolution of the beamformed dynamic spectra are 1 MHz and 1 \textmu s respectively. Delay, amplitude, and band-pass gain calibration solutions are applied to both the imaging and beamformed data products, while the polarisation calibration solutions are derived from beamformed observations of the polarisation calibrator, and applied to the beamformed dynamic spectra of the FRB. Depending on the local sidereal time at the time of the calibration observation, we utilised either the Vela pulsar (J0835-4510) or PSR J1644-4559 to provide polarisation calibration. While Vela is an ideal calibrator across the entire ASKAP observing band, PSR J1644-4559 becomes more highly scattered and less highly linearly polarised at lower frequencies, and provides lower quality polarisation calibration solutions in the low ASKAP band (below 1 GHz). PSR J1644-4559 was observed as a polarisation calibrator for four low band FRBs and therefore we have not applied polarisation calibrator solutions to these bursts. Bursts with calibration solutions applied are denoted by a green star in Fig. \ref{fig:dynspec_mosaic}. Despite the lack of polarisation calibration for the remaining bursts, we expect any errors in the polarisation observables to be $\lesssim$5 per cent \citep{glowacki2026pinkupdateimprovementscelebi}.

De-dispersion is performed based on a structure-maximization routine described by \cite{sutinjo2023calculation} and implemented in the structure maximisation of high-time resolution intensity profiles with no-nonsense errors (\textsc{SHRINE}\footnote{https://github.com/marcinglowacki/SHRINE}) codebase. The structure-maximized DM is applied to the voltages before re-constructing optimal Stokes dynamic spectra. We summarise the best-fit DM values for each burst in Table \ref{tab:burstprop}. 

The latest CRAFT FRB sample consists of 35 bursts from the previous CRAFT data release \citep{scott2025high} and the 34 bursts presented here. For this analysis we will only study the `apparent' non-repeating FRBs. Thus, we remove the confirmed repeater burst FRB 20190711A from the \citet{scott2025high} sample. We refer to the 34 bursts from the prior data release as \oldhtr{}. Similarly, we remove the two repeat bursts from FRB 20250613A \citep{dial2026} from our latest sample. Furthermore, we exclude FRBs 20260217D and 20260307D since they have incomplete voltage downloads that provide only a single linear polarisation. This leaves 30 bursts in the latest sample of CRAFT FRBs which we refer to as \newhtr{}; the four excluded bursts from \newhtr{} are included as spectrograms in Fig.~\ref{fig:dynspec_mosaic} for completeness. For the remainder of this analysis we study the polarisation properties of 64 CRAFT bursts and search for any potential correlation with burst energetics.

\subsection{Analysis}

The following analysis is performed on the structure-maximized, coherently beamformed Stokes dynamic spectra of each burst in \newhtr. The dynamic spectra has an intrinsic time and frequency resolution of 1 \textmu s and 1 MHz respectively. Fig.~\ref{fig:dynspec_mosaic} shows the dynamic spectra of these bursts; for the purposes of visual clarity the dynamic spectra were downsampled in both time and frequency in order to balance S/N and the ability to resolve fine temporal structure, and the time and frequency resolution are annotated on each panel in Fig.~\ref{fig:dynspec_mosaic}.

\subsubsection{RFI flagging and width}
\label{sec:widthfitting}

To remove channels afflicted by strong RFI, we use the statistical approach detailed in \cite{dial2026}. Channels are removed when the standard deviation exceeds the median of the full bandwidth by a threshold $M$:

\begin{equation}
    \Big|\sigma_{t}(f) - \mathrm{med}\big(\sigma_{t}(f)\big)\Big|
        > M  \cdot \mathrm{med}\Big(\Big|\sigma_{t}(f) - \mathrm{med}\big(\sigma_{t}(f)\big)\Big|\Big)
\end{equation}

where $\sigma_{t}(f)$ = $\sqrt{\sum_{t}I(f,t)^2}$. For each burst, the Stokes $I$ dynamic spectrum is down-sampled to a resolution of 1 ms before performing RFI flagging.

Again following \cite{dial2026}, we measure the on-pulse region of each burst in both time and frequency. Briefly, we scrunch the Stokes \textit{I} dynamic spectrum in frequency and look for the minimum width of the burst that incorporates at least 95\% of the total fluence, which we denote as $w_{95}$. Compared to time series fitting approaches such as those described in \cite{dial2025frb} and \cite{dial2026}, which model an intrinsic burst profile (generally described by one or more components) that is convolved with a pulse broadening function, this is far simpler and faster. However, it does not provide an estimate of either the intrinsic burst width or the temporal broadening (scattering), which we defer to a future work. We apply this width as a boxcar filter and scrunch the Stokes \textit{I} dynamic spectrum in time to create a 1D spectrum. Summing in time, we look for the minimum bandwidth of the burst that incorporates at least 95\% of the total fluence, which we denote as $\Delta \nu_{95}$. 

We perform this two-step process twice to optimize both $w_{95}$ and $\Delta \nu_{95}$, which we include in Table \ref{tab:burstprop}.

\begin{figure*}
    \centering
    \includegraphics[width=0.87\textwidth]{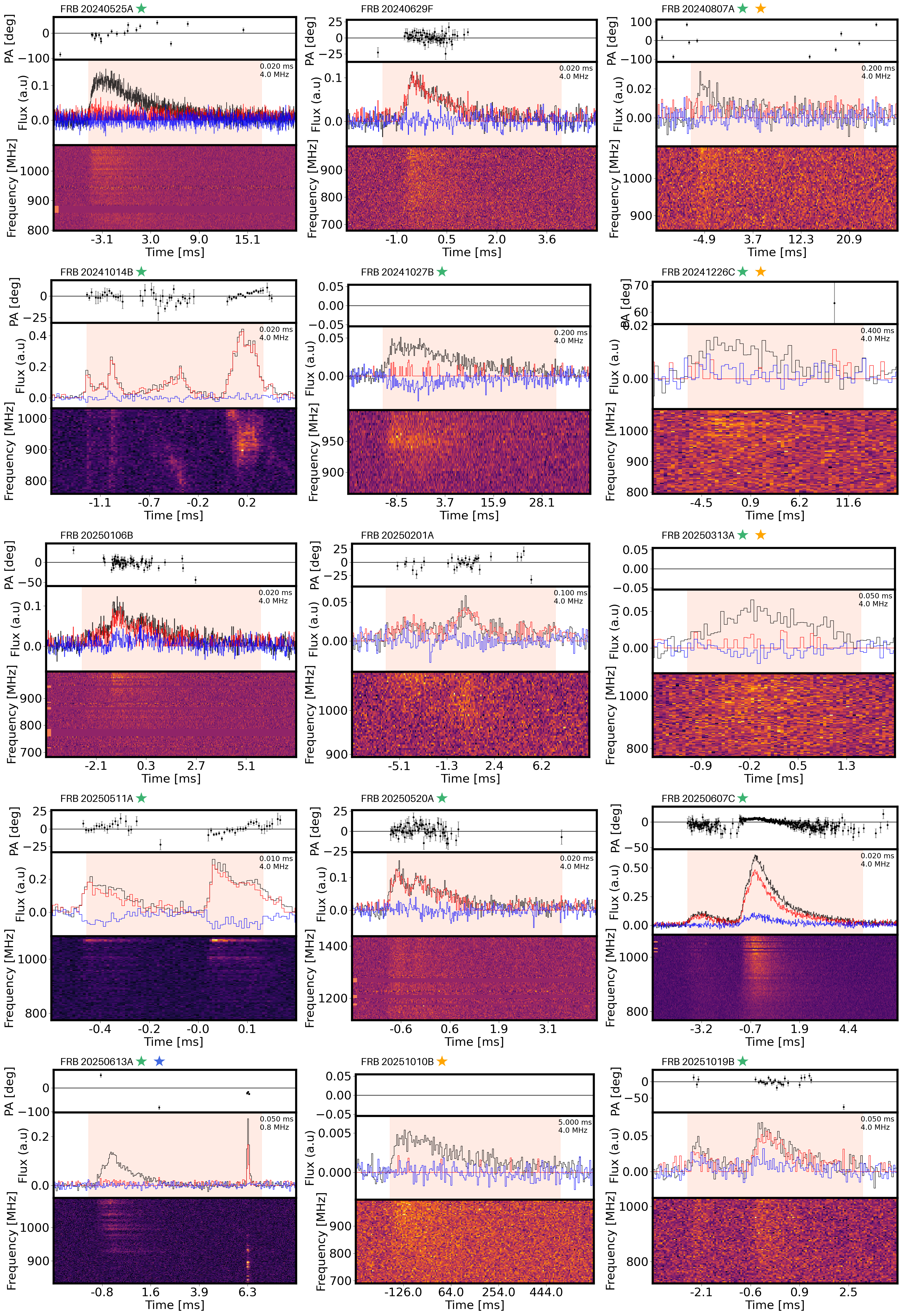}
    \caption{Mosaic of CRAFT FRBs, dedispersed to the structure-maximising DM and averaged in time and frequency for visual clarity. Top panel: PA profile. Middle panel: Stokes \textit{I, L} and \textit{V} time series, time and frequency resolution are labeled in the upper right corner of the panel. Bottom panel: Dynamic spectra. Flagged channels are denoted by orange markings. Green star next to FRB name indicates the burst has been properly polarisation calibrated. Orange star denotes the burst is classified as unpolarised. Blue stars show bursts from repeaters, specifically FRB 20250613A.}
    \label{fig:dynspec_mosaic}
\end{figure*}

\begin{figure*}
    \ContinuedFloat
    \centering
    \includegraphics[width=0.9\textwidth]{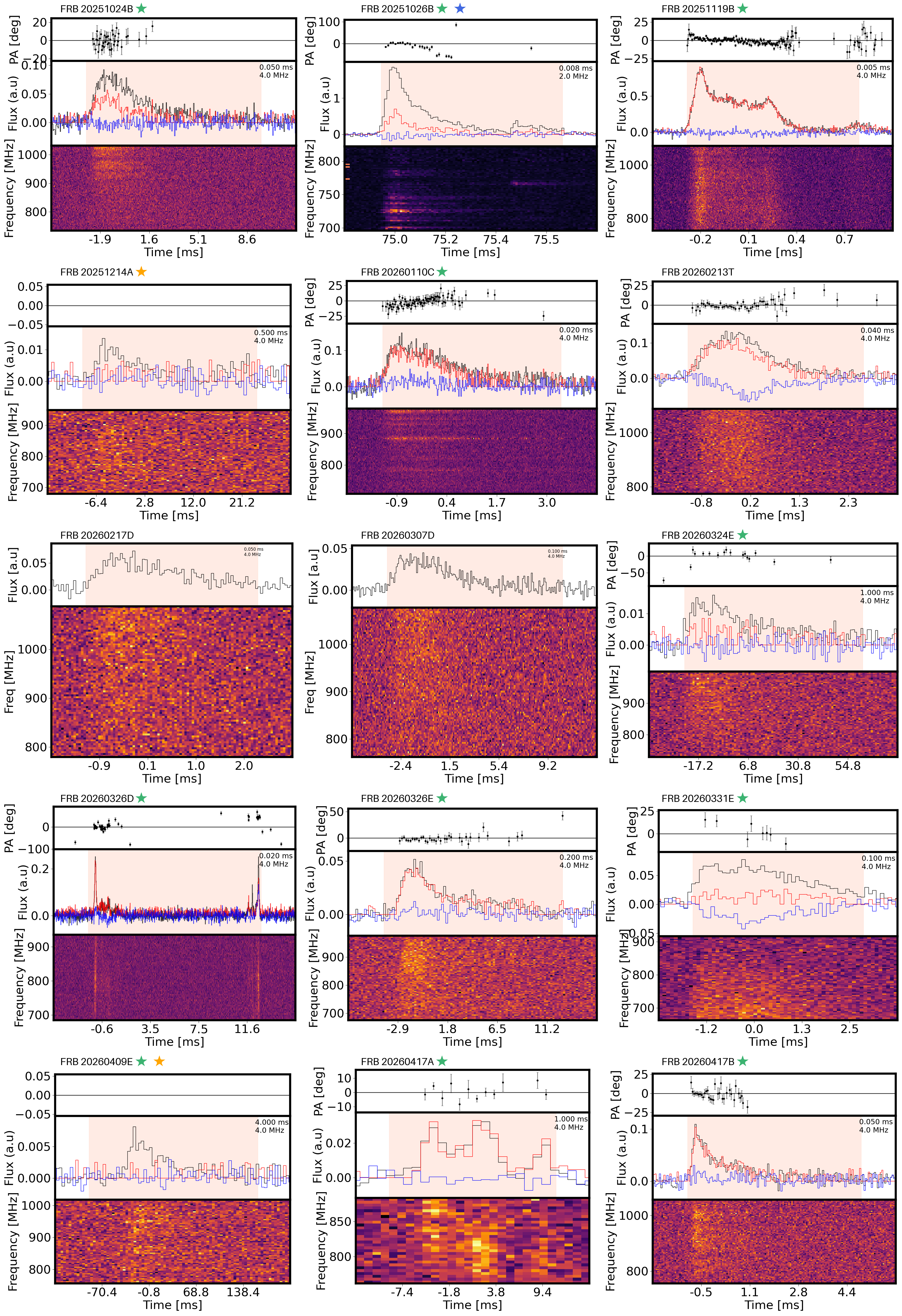}
    \caption{(cont.)}
\end{figure*}

\begin{figure*}
    \ContinuedFloat
    \centering
    \includegraphics[width=0.9\textwidth]{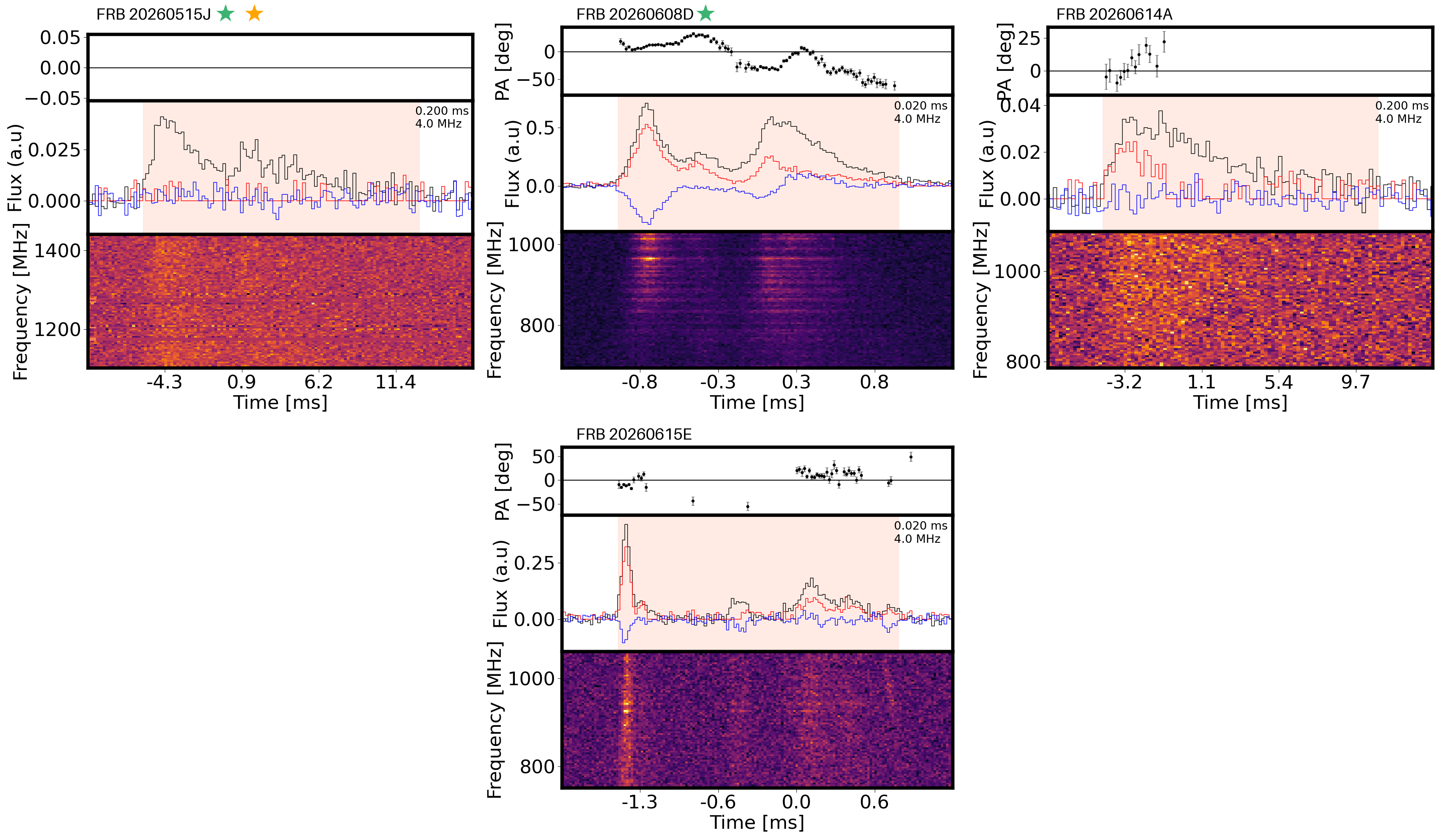}
    \caption{(cont.)}
\end{figure*}

\subsubsection{RM fitting}
\label{sec:RMfitting}
 A boxcar filter with a width of $w_{95}$ was applied to the Stokes \textit{I}, \textit{Q} and \textit{U} dynamic spectrum to produce a one-dimensional spectrum. We derived the rotation measure for each burst by processing the one-dimensional spectra using the RM synthesis \citep{brentjens2005faraday} python package \textsc{RMtools}. The RM and its uncertainty along with the power-averaged central frequency $\nu_{0}$ are recorded in Table \ref{tab:burstpol}.

We corrected the Faraday rotation in the Stokes \textit{Q} and \textit{U} dynamic spectra by applying a rotation in \textit{Q}-\textit{U} space:

\begin{equation}
    \label{eq:FdayRot}
    \begin{split}
        & Q\mathrm{_{deRM}} = Q\mathrm{cos(2PA)} + U\mathrm{sin(2PA)} \\
        & U\mathrm{_{deRM}} = Q\mathrm{sin(2PA)} - U\mathrm{cos(2PA)}, \\
    \end{split}
\end{equation}

where $\mathrm{PA}$ is the polarization position angle:

\begin{equation}    
    \label{eq:RM}
    \mathrm{PA(\nu) = RM\,c^{2}}\bigg(\frac{1}{\nu^{2}} - \frac{1}{\nu_{0}^{2}}\bigg).
\end{equation}

 We searched over a grid of RM values out to a maximum |RM| of 10$^{4}$ rad~m$^{-2}$. If no significant peak was found in this range we extended the search to a maximum |RM| of 10$^{5}$ rad~m$^{-2}$ after re-channelising the voltage data to a higher channel resolution (as fine as 50 kHz for low-band bursts) and re-forming the dynamic spectra. For seven of the 30 non-repeating bursts, we did not find a significant peak in this range, meaning we could not identify a RM and can only place an upper limit on the polarisation fraction. We refer to these bursts as unpolarised and have denoted them with orange stars in the gallery of dynamic spectra in Fig. \ref{fig:dynspec_mosaic}.

We estimate the Galactic RM contribution using the Spectra and Polarisation in Cutouts of Extragalactic Sources \citep[SPICE-RACS;][]{thomson2026rapid} second data release, which is a densely sampled reconstruction of the Galactic Faraday sky which has a typical median uncertainty of $\sim$2 rad~m$^{-2}$. The estimated Galactic RM was subtracted from the measured RM of each burst to obtain the extra-galactic contribution:

\begin{equation}
    \rm RM_{\mathrm{EG}} = RM - RM_{\mathrm{MW}}.
\end{equation}

Both the RM and Galactic RM are reported in Table~\ref{tab:burstpol}. We also updated the Galactic RMs for the \oldhtr{} burst sample which are reported in Table~\ref{tab:htr1_prop}.

\subsubsection{Polarisation position angle and fractions}

The measured PA profile of each burst $\psi$ is given by:

\begin{equation}
    \psi(t) = \frac{1}{2}\arctan\bigg(\frac{U_{\mathrm{deRM}}(t)}{Q_{\mathrm{deRM}}(t)}\bigg).
\end{equation}

$\psi$ and the uncertainty $\sigma_{\psi}$ are de-biased following \cite{day2020high}. The PA is plotted for each burst in Fig.\ref{fig:dynspec_mosaic}.

The frequency-averaged one-dimensional Stokes \textit{I, Q, U} and \textit{V} time series profiles are used to calculate the linear (\textit{L/I}), absolute circular (\textit{|V|/I}) and total (\textit{P/I}) polarisation fractions for each burst:

\begin{equation}
    \begin{split}
        &x = \frac{\sum_tX(t)}{\sum_tI(t)}\hspace{0.1cm}\mathrm{for}\hspace{0.1cm}x\in \{q,u,|v|,l\} \hspace{0.1cm} \rm and \hspace{0.1cm}X\in \{Q,U,|V|,L\},\\
        &\mathrm{where}\\
        &L = \sqrt{Q^2 + U^2}. \\  
    \end{split}
\end{equation}

We calculate $p$ in the same manner as the previous HTR data release \citep{scott2025high}:

\begin{equation}
    p = \sqrt{l^2 + |v|^2}.
\end{equation}

Stokes \textit{L} and \textit{P} are de-biased following \cite{day2020high} and \cite{mckinnon2025polarization} whilst the magnitude of Stokes \textit{V} (\textit{|V|}) is de-biased according to \cite{karastergiou2003v, oswald2023pulsar1, mckinnon2025polarization}. All polarisation fractions are reported in Table \ref{tab:burstpol}.

\subsubsection{Fluence and isotropic spectral luminosity}

Transfer of the amplitude calibration derived in the CELEBI pipeline to the beamformed data is complicated by the normalisation of the polyphase filterbank inversion used to obtain the highest possible time resolution for the beamfomed data \citep{scott2023celebi}. Instead of employing the amplitude calibration from the uninverted cross-correlation data, we simply estimate the burst fluence using the average antenna system equivalent flux density (SEFD) given by \cite{hotan2021australian}:

\begin{equation}
    \mathrm{SEFD} = \frac{2k_{b}}{A}\frac{1}{N_{\rm chan}}\sum^{N_{\rm chan}}\frac{T_{\mathrm{sys}}(f)}{\eta},
\end{equation}

where $A$ is the geometric area of a single dish, $k_{b}$ is the Boltzmann constant and the factor of 2 accounts for the two polarisation receptors at the receivers. The effective system temperature $T_{\mathrm{sys}}(f)/\eta$ is modelled as a polynomial based on Fig.~22 of \cite{hotan2021australian} and averaged over the burst bandwidth, where $N_{\rm chan}$ is the number of channels. The burst fluence $F$ (measured in Jy ms) is calculated using the SEFD:

\begin{equation}
    F = C\frac{\mathrm{SEFD}}{N_{\mathrm{ant}}B}\sqrt{\frac{w_{95}}{\Delta\nu_{95}}}(S/N)_{\rm int},
\end{equation}

where $N_{\mathrm{ant}}$ is the number of antenna, $w_{95}$ is the burst width in milliseconds, $\Delta\nu_{95}$ is the burst bandwidth in MHz, $(S/N)_{int}$ is the integrated S/N, $B$ represents the response of the downloaded ASKAP PAF beam at the location of the FRB, normalised relative to the PAF beam peak, and $C$ = $\sqrt{1\times10^{3} \rm [ms~s^{-1}]}/\sqrt{1\times10^{6} \rm [Hz~MHz^{-1}]}$ = 0.0316 is a combination of unit conversion scalars for the width and bandwidth. We model the PAF beam response $B$ as a gaussian function \citep{condon2016essential}:

\begin{equation}
    B = \rm exp\bigg[-4ln(2)\bigg(\frac{\theta_{b}}{\theta_{FWHM}}\bigg)^{2}\bigg],
\end{equation}

where $\rm\theta_{FWHM}$ = 1.22$\lambda$/D is the gaussian beam full width at half maximum and $\theta_{b}$ is the beam angular offset from boresight in radians. $\lambda$ is the central wavelength of the observation and D = 12 m is the diameter for an ASKAP dish. Using the fluence we calculate the isotropic spectral luminosity:

\begin{equation}
\label{eq:speclum}
    L_s = 1.257\times10^{-18}\frac{Fd_{L}^{2}}{w(1+z)},
\end{equation}

where $d_{L}$ is the luminosity distance which is estimated assuming a $\Lambda$-CDM cosmology with a Hubble constant of $H_{\mathrm{0}}$ = 67.4 km s$^{-1}$ Mpc$^{-1}$ and a matter density of $\Omega_{m}$ = 0.315 \citep{aghanim2020planck}. For the analysis and discussion of this paper, we also measured the fluence and isotropic spectral luminosity for each of the bursts in the \oldhtr{} sample.

\section{Discussion}
\label{sec:discussion}

\subsection{Polarisation fractions}
\label{sec:pcorr}

\subsubsection{Comparison with CHIME and DSA}

We studied the distribution of L/I and V/I for the latest CRAFT sample of 64 bursts and compared them with the CHIME and DSA burst samples. The CRAFT sample consists of 57 bursts with measured polarisation properties (34 from \oldhtr{} and 23 from \newhtr{}) and seven newly discovered unpolarised bursts. We include 89 polarised bursts and 29 unpolarised bursts from the CHIME sample \citep{pandhi2024polarization}. Similarly, we include 20 polarised bursts and five unpolarised from the DSA sample \citep{sherman2024deep}. For the seven unpolarised CRAFT FRBs we estimate an upper limit on L/I by adopting a S/N threshold of 6 \cite[i.e S/N$_{L}\ge$~6.0;][]{pandhi2024polarization}. This detection threshold is more conservative as it is slightly above the minimum S/N$_{L}$ of 5.0 needed to obtain a reliable RM measurement using \texttt{RM-Tools} \citep{fine2023correcting, van2026rm}. Because V/I is not affected by the RM, we use a detection threshold of S/N$_{V}\ge$ 3.0. We also place an upper limit on L/I for FRB 20241027B, where we were unable to measure an RM but which has a high V/I of $\sim$65\%. Since each sample consists of a sub-set of unpolarised bursts with upper limits on fractional polarisation we used a Kaplan-Meier estimator \citep{kaplan1958nonparametric} to determine the cumulative distribution function (CDF) by treating the unpolarised bursts as left-censored measurements through the \textsc{lifelines} survival analysis python package \citep{Davidson-Pilon2019}. Bootstrapping with replacement was used to determine the median and 95\% confidence intervals for each CDF.

\begin{figure}
    \centering
    \includegraphics[width=\linewidth]{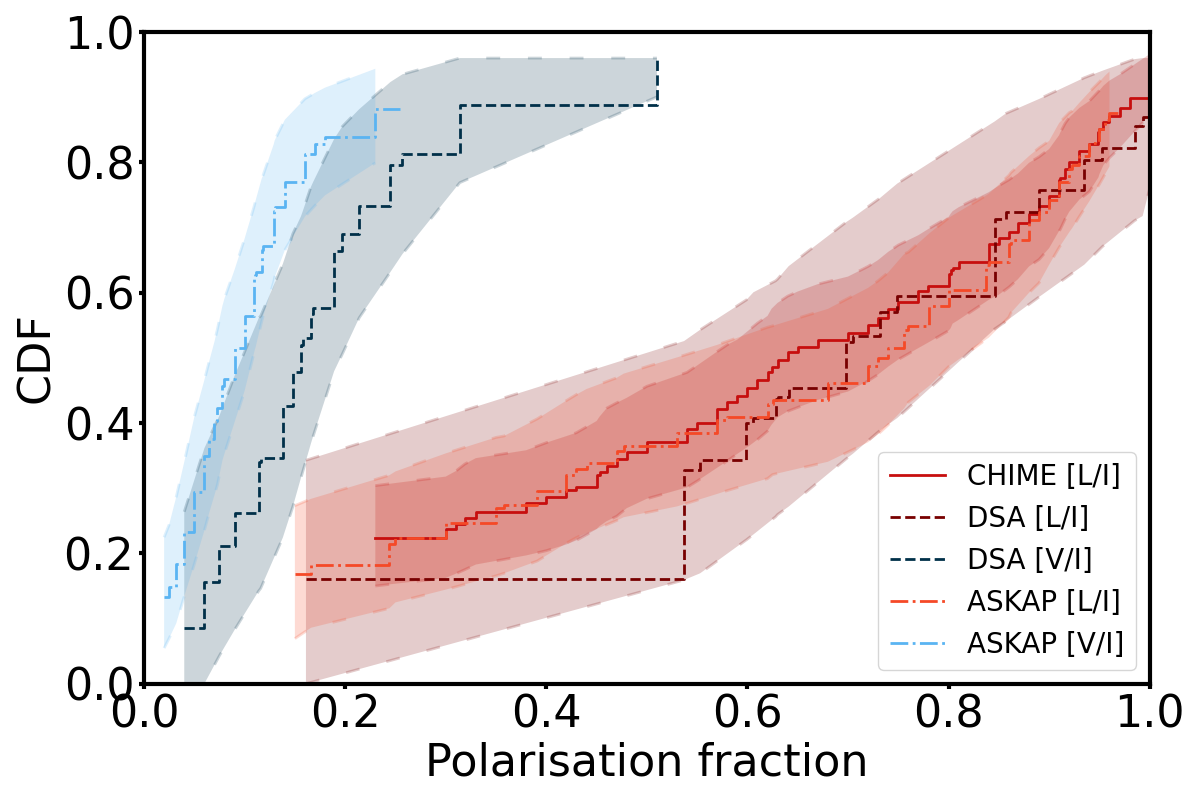}
    \caption{Cumulative distribution of linear and circular polarisation fraction. The bold lines show the median CDF whilst the shaded regions show the 95\% confidence intervals.}
    \label{fig:polfrac_samples}
\end{figure}

In Fig. \ref{fig:polfrac_samples} we show the L/I and V/I CDFs for the ASKAP and DSA samples and the L/I CDF for the CHIME sample (V/I is not reported by CHIME due to the difficulty in modelling the complex direction- and frequency-dependent polarisation response of the CHIME telescope). In \cite{scott2025high} the \oldhtr{} sample of ASKAP bursts generally exhibited higher fractions of L/I than CHIME and DSA. However, with the addition of the \newhtr{} bursts there are very minor qualitative difference between ASKAP and both CHIME and DSA. We use a log rank statistic \citep{peto1972asymptotically} to quantify the difference between the three major FRB samples using Monte Carlo simulations. Comparing the L/I distribution between ASKAP and CHIME we obtain a test statistic of 2.32 and an associated $p$-value of 0.13. Between ASKAP and DSA the test statistic and $p$-value is 0.64 and 0.42 respectively. There is therefore no evidence supporting a difference in the underlying L/I distribution seen by any of the three surveys. 


Despite the consistency in linear polarisation fraction, however, DSA shows strong evidence for a higher average V/I distribution compared to ASKAP with a test statistic and $p$-value of 7.63 and 0.008 respectively. One potential explanation would be the differences in the polarisation purity and calibration techniques used. The polarisation calibration techniques employed to calibrate ASKAP FRBs are described in detail by \cite{glowacki2026pinkupdateimprovementscelebi} and involve correcting for two primary sources that affect the true polarimetry of the signal: (1) Spectral leakage due to imperfections between the digitisers that process the two linear polarisation modes at the antenna feed which cause phase delays between the incoming modes and are seen as a rotation between \textit{U} and \textit{V} in Stokes parameter space; and (2) imperfections in the linear polarisation feeds that cause coupling between the polarisation modes, thus appearing generally elliptical when incident on the antenna feed. This is seen as a rotation between Stokes \textit{Q} and \textit{V}. Both sources result in reducing L/I and increasing V/I. For most ASKAP bursts the mode coupling is negligible, only causing significant leakage ($>$5\%) in cases where the burst is incident on a corder/edge beam in the phased array feed (PAF) on the antenna \citep{glowacki2026pinkupdateimprovementscelebi}. Within our sample, only one burst (FRB 20210320C) meets these conditions, meaning the beam shape plays an effect on its polarisation. The spectral leakage present in the digitisers is also small in most FRBs, and only introduces a small error if polarisation calibration is not performed. Thus, while we do not have an explanation for the differences seen in the V/I distributions between ASKAP and DSA, we are confident in the accuracy of the ASKAP polarisation measurements. 

\subsubsection{Correlation with other burst properties}
\begin{figure*}
    \centering
    \includegraphics[width=\textwidth]{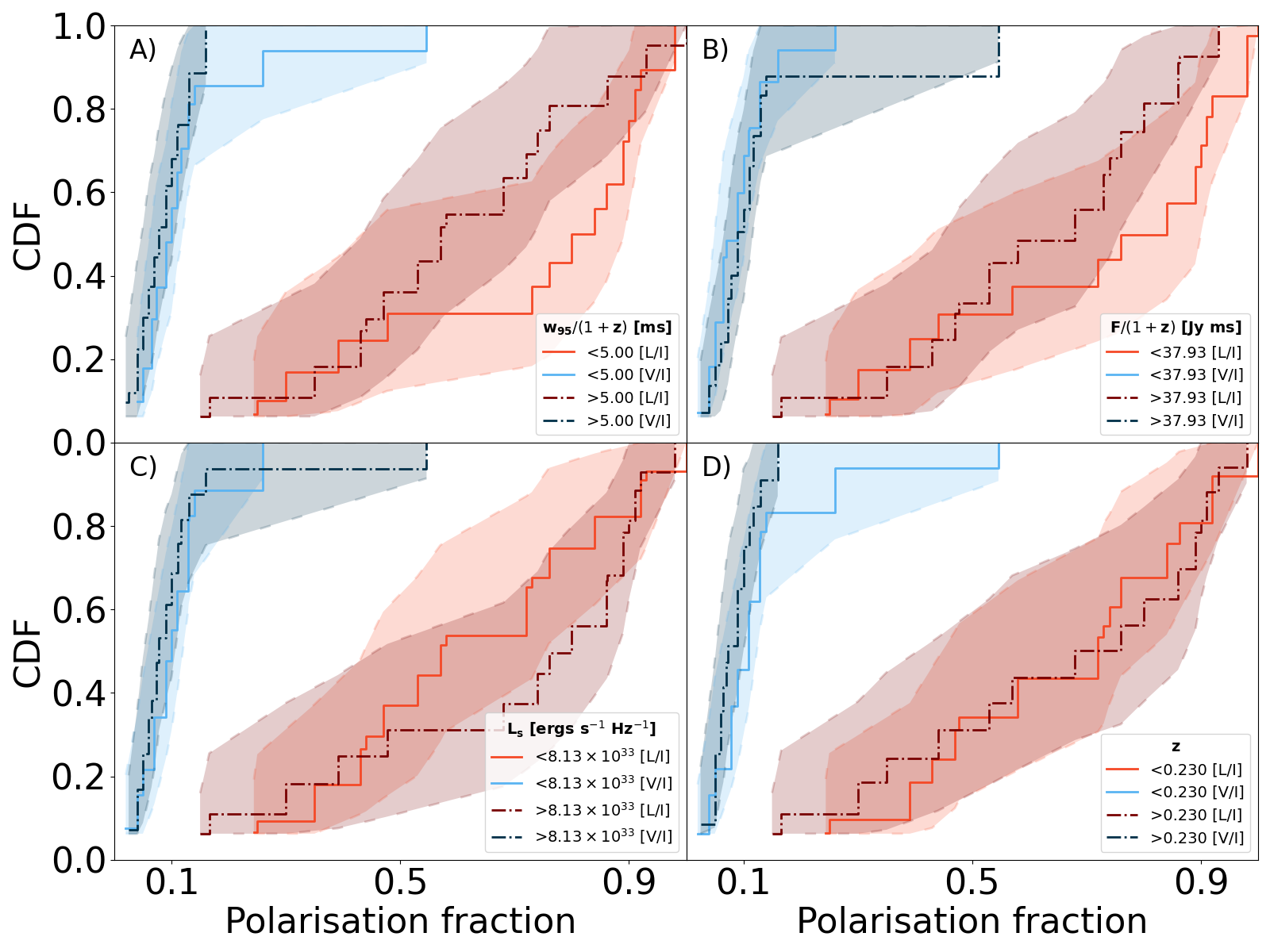}
    \caption{L/I and V/I CDFs after splitting ASKAP FRBs by width (panel A), fluence (panel B), spectral luminosity (panel C) and redshift (panel D). Bold lines represent the median CDF and the shaded region shows the 95\% confidence interval.}
    \label{fig:LV_energ}
\end{figure*}

We investigate potential correlations between the observed polarisation fractions and other burst properties, including apparent width, fluence and average spectral luminosity. When calculating luminosity, we consider only sources with a reliable distance given by a confident host galaxy association and a spectroscopic redshift (see Table~\ref{tab:burstprop} and the references provided). Additionally, we removed four multi-component bursts whose sub-component separations were comparable to the estimated burst width $w_{95}$ (e.g. FRB 20260326D), since in these cases, the average spectral luminosity differs significantly from the on-pulse luminosity. Lastly, we removed FRB 20190102C from \citet{scott2025high} because the burst was detected at a large angular offset (2.5 deg) from boresight where the primary beam corrections introduce poorly constrained systematic errors. After filtering, 32 bursts remained. These bursts mostly come from the \oldhtr{} sample since more time has passed to obtain redshift measurements, meaning that relatively few of the (on average wider and fainter) CRACO bursts are included. We corrected the observed width and fluence with a redshift factor of 1/(1+$z$) and measured the isotropic spectral luminosity using equation \ref{eq:speclum}. We divide the sample into two evenly distributed bins by their width (using the median width in the sample) and for each bin estimated the L/I and V/I CDFs as shown in panel A of Fig. \ref{fig:LV_energ}. We did not use the Kaplan-Meier estimator since we did not include the unpolarised bursts. We instead used a two-sample Anderson-Darling test \citep{anderson1952asymptotic} to estimate the test statistic and associated $p$-value which are reported in Table~\ref{tab:statistics}. We repeated this process three more times, splitting the sample by fluence, spectral luminosity and redshift. 


\begin{table}
\centering
\setlength\tabcolsep{12pt}  
    \caption{Derived test statistic and $p$-value using the two-sample Anderson-Darling test and comparing against both the L/I and V/I sub-sample distributions after splitting the ASKAP burst sample by width, fluence, spectral luminosity and redshift. } 
    \label{tab:statistics}
    \begin{tabular}{ccccc}
    \hline
         &  \multicolumn{2}{c}{Test statistic}  &     \multicolumn{2}{c}{$p$-value} \\ \hline
    Parameter    &   L/I     &   V/I     & L/I       & V/I \\ \hline
    Width    &    0.66    &   -0.49     &    0.18   &  0.62 \\
    Fluence    &  0.59      &   -0.50     &   0.20      &  0.62\\
    Spectral Luminosity    &   -0.14     &    -0.79    &    0.42   &  0.82\\
    Redshift    &    -0.96    &    0.02    &   0.94     &  0.36 \\ \hline
    \end{tabular}
\end{table}

In none of the examined properties is there a significant difference seen in the polarisation fraction between the sub-divided populations. The largest differences are seen when dividing the data by the total width (see panel A of Fig.~\ref{fig:LV_energ}), with narrower bursts on average possessing higher fractions of L/I, but the associated $p$-value of 0.18 is not statistically significant, motivating future studies with a larger sample. Effects such as multipath propagation could in principle lead to depolarisation along with temporal broadening, which would result in the observed tendency for narrower bursts to be more highly polarised. The intrinsic width and scattering times of the latest CRAFT sample will be extracted in future work, allowing the effects to be examined separately. 

When dividing the data by fluence (or spectral luminosity), lower fluence (or more luminous) bursts exhibit higher L/I values. However, these differences are smaller compared to those seen when dividing by burst width and, unsurprisingly, show no statistical significance, with $p$-values $>$0.2. Nevertheless, in the event that burst width does play a role in the observed polarisation fraction, we would also expect some dependence on both fluence and luminosity; a larger sample in the future is required to ascertain whether this trend persists at a statistically significant level. 

We note that there is no evidence to support any of these observed properties having an effect on the V/I fraction, given the measured $p$-values $>$0.35. Although qualitatively there seems to be a sharp divergence at V/I $>$0.2, the tail of this distribution is dominated by FRB 20260608D which has a large |V|/I of 57\%. FRB 20260608D also has a very large apparent brightness (i.e., a large S/N; see Fig.~\ref{fig:dynspec_mosaic}), which strongly skews the V/I distribution in fluence, as seen in panel B of Fig.~\ref{fig:LV_energ}. Therefore, V/I appears to show no true dependence on these properties. 

\begin{figure}
    \centering
    \subfigure[]{\includegraphics[width=\linewidth]{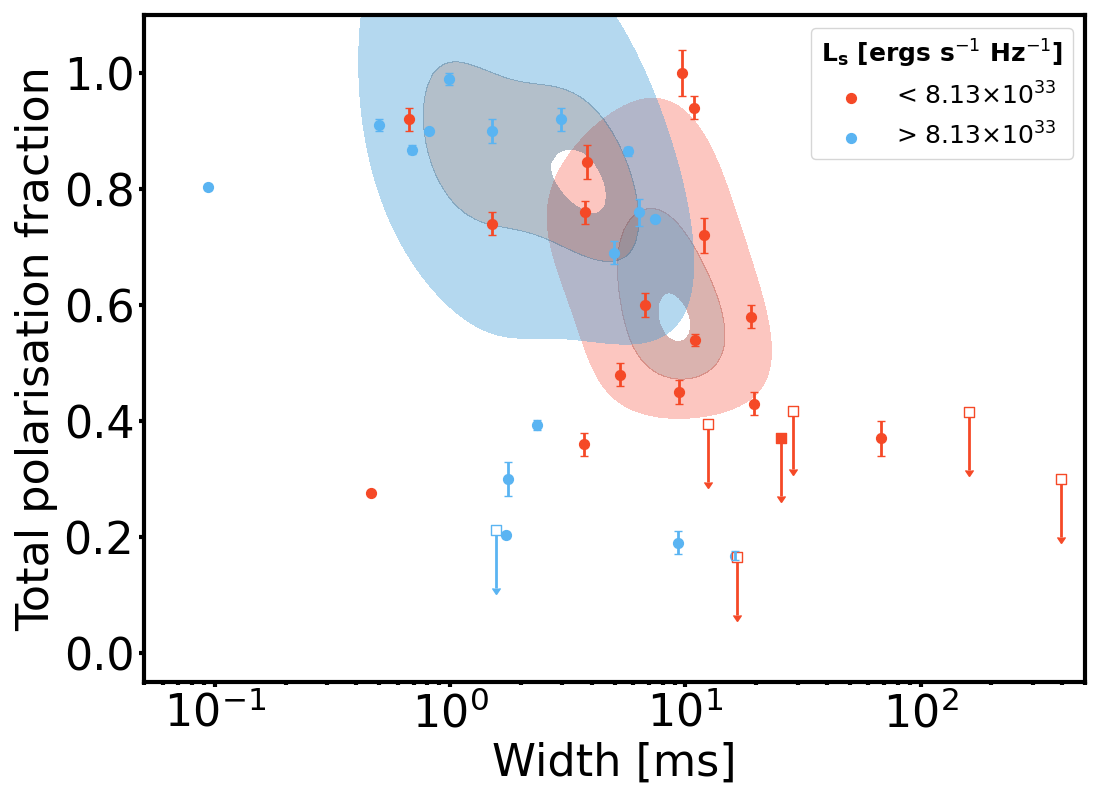}}
    \subfigure[]{\includegraphics[width=\linewidth]{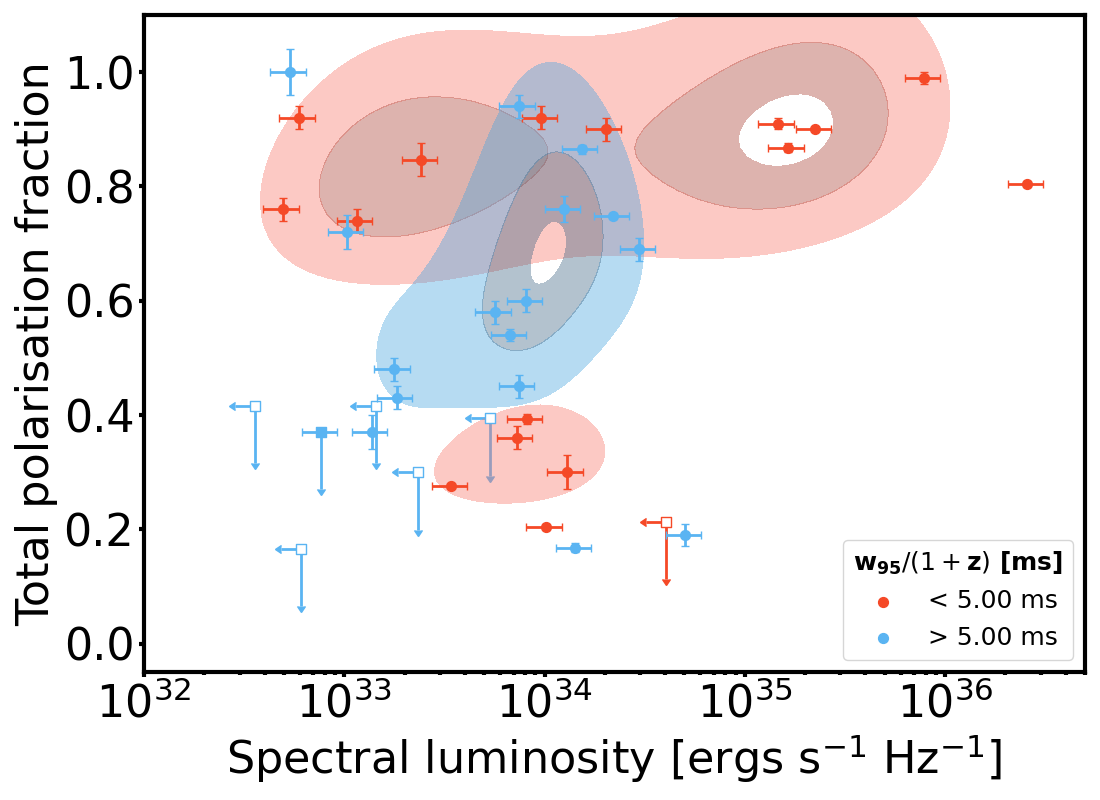}}
    \caption{(a) Probability distribution of ASKAP sample as a function of $p$ and width after splitting by spectral luminosity. (b) Probability distribution of ASKAP sample as a function of total polarisation fraction $p$ and spectral luminosity after splitting by width. Contour lines show the 68, 90 and 99 per cent confidence intervals. Circle markers show polarised bursts and square markers show unpolarised bursts. Solid square marker shows unpolarised burst with confirmed redshift measurement. For six out of the seven unpolarised bursts, the redshift was estimated from the extragalactic DM as described in the text. These bursts are represented as hollow square markers.}
    \label{fig:pVenerg}
\end{figure}

We show the two-dimensional probability distribution of P/I against width after binning the data by spectral luminosity in panel (a) of Fig. \ref{fig:pVenerg}. This shows the characteristics of the individual bursts driving the observed differences in the ASKAP population to date: very wide bursts ($\gtrsim$20 ms) are invariably of lower luminosity (unsurprising given their width) and have low levels of linear polarisation (panel (a) of Fig.~\ref{fig:pVenerg}). Many of these bursts have no detected polarisation which we explore in Section~\ref{sec:unpol}.  Conversely, highly energetic bursts ($L_{s}>$10$^{35}$ ergs~s$^{-1}$~Hz$^{-1}$) are consistently both narrow and highly linearly polarised (panel (b) of Fig.~\ref{fig:pVenerg}).  
Bursts that are not extremely wide can be either bright or faint (showing high or low polarisation fractions), and bursts that are not extremely luminous can similarly be narrow or wide, but bursts at the high extremes of the width and luminosity distribution strongly prefer low and high polarisation fractions respectively.

\subsection{Un-polarised bursts}
\label{sec:unpol}

Seven out of the 30 bursts analysed in this work with full baseband data are classified as unpolarised, with a typical 6.0 sigma upper limit on the polarisation fraction of 37\%. Interestingly, ASKAP only started uncovering this population of unpolarised bursts after the commissioning of CRACO, which provides higher sensitivity and enables the detection of fainter (and wider, due to the lower time resolution) bursts. A subset of bursts from both the DSA and CHIME FRB catalogs -- which, like CRACO, have higher sensitivity compared to the ASKAP ICS mode -- are also unpolarised \citep{sherman2024deep, pandhi2024polarization}. This suggests the population of unpolarised bursts are typically fainter. 
Unfortunately, as most of these unpolarised CRACO bursts were detected relatively recently, just one of the seven unpolarised bursts, FRB 20240807A, has a measured redshift of $z$ = 0.2097, from which we calculate a redshift corrected width of 25.46 ms and an isotropic spectral luminosity of 7.67$\times$10$^{32}$ ergs s$^{-1}$ Hz$^{-1}$, making FRB 20240807A one of the widest and faintest bursts in the ASKAP sample. For six out of the seven bursts no spectroscopic redshift is available at this time. For these, we estimate a redshift based on the extragalactic DM, DM$_{\mathrm{EG}}$, which is measured by subtracting the DM contributions from the Milky Way interstellar medium (DM$_{\mathrm{MW,ISM}}$) and halo (DM$_{\mathrm{MW, Halo}}$). We use the NE2025 Galactic electron density model \citep{ocker2026ne2025} to calculate DM$_{\mathrm{MW, ISM}}$. DM$_{\mathrm{MW,Halo}}$ has been shown to vary drastically over a range of different Galactic latitudes with estimates between 40-120 pc cm$^{-3}$ \citep{prochaska2019probing, keating2020exploring, Cook_2023, ravi2025}. A recent study by \cite{hoffman2026dm} reports a halo contribution of DM$_{\mathrm{MW, Halo}}$ = 68$^{+27}_{-24}$ pc cm$^{-3}$ for Galactic latitudes above 20$^{\circ}$. Here we use a nominal value of DM$_{\mathrm{MW,Halo}}$ = 50 pc cm$^{-3}$ and note that changes to this contribution do not materially change the results that follow. Accounting for the Galactic contributions, DM$_{\mathrm{EG}}$ becomes a function of redshift:

\begin{equation}
\label{eq:DMz}
    \mathrm{DM_{EG}}\text{(}z\text{)} = \mathrm{DM_{IGM}}z + \frac{\mathrm{DM_{Host}}}{1+z},
\end{equation}
where DM$_{\mathrm{Host}}/(1+z)$ is the rest frame host DM. We perform a linear fit of DM$_{\mathrm{EG}}$ against redshift to get DM$_{\mathrm{IGM}}$ using equation 14 from \cite{macquart2020census} assuming cosmological parameters from \cite{aghanim2020planck}. We estimate redshift by rearranging equation \ref{eq:DMz}. 

The sample of unpolarised bursts are shown in Fig.~\ref{fig:pVenerg}. On average, unpolarised bursts are significantly wider and exhibit lower spectral luminosities. This corroborates our findings in Section~\ref{sec:pcorr} that fainter, wider bursts possess lower polarisation fractions. Future samples of fainter, wider FRBs - such as those discovered by CRACO - will be instrumental in uncovering the interplay between emission mechanism and propagation effects that influence both width and luminosity.


\subsection{Spectral depolarisation}

A number of repeating FRBs exhibit a signature of spectral depolarisaton, where the polarisation fraction decreases at lower frequencies \citep[e.g., FRB 20201124A; ][]{lu2023temporal}. This has been modelled as resulting from multi-path propagation through a highly turbulent, magneto-ionised local environment \citep{burn1966depolarization, feng2022frequency}. Amongst the small sample of repeating FRBs detected by CRAFT, FRB 20250613A was discovered by ASKAP as part of the \newhtr{} sample and exhibits clear spectral depolarisation \citep{dial2026}. In contrast, spectral depolarisation has been rarely reported for apparent non-repeating FRBs to date, suggesting these sources typically reside in less complex environments. However, recent work by \cite{uttarkar2024, 2026MNRAS.545f1997U} examined a total of 17 CRAFT non-repeating FRBs and found evidence for spectral depolarisation in FRB 20230526A. The polarimetry study of 25 DSA FRBs by \cite{sherman2024deep} found three bursts that could be replicated with minor spectral depolarisation between 0.3$\leq\mathrm{\sigma_{RM}}\leq$3.0 rad~m$^{-2}$, where $\mathrm{\sigma_{RM}}$ characterises the strength of the turbulence in the intervening plasma. 

\begin{figure}
    \centering
    \includegraphics[width=\linewidth]{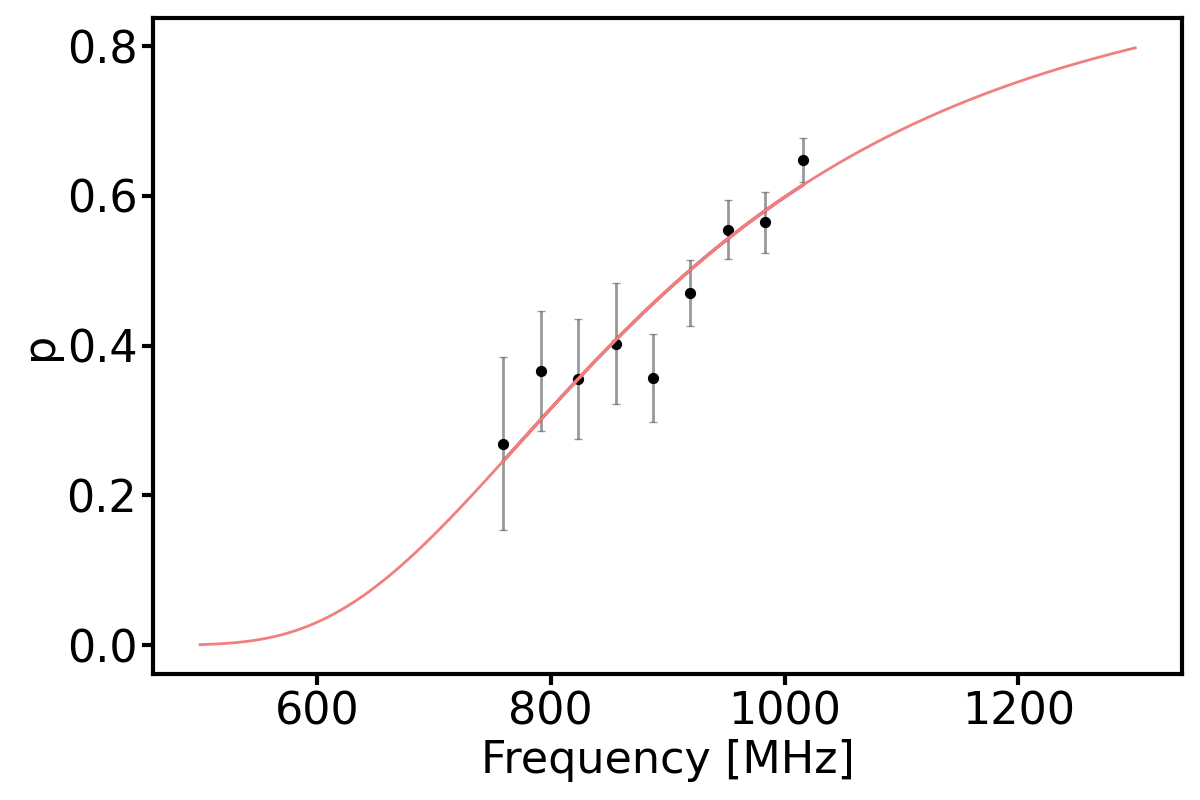}
    \caption{Spectral depolarisation modelling of FRB 20251024B. Red line represents the best fit using the modified burns law (equation \ref{eq: burnslaw}) with an intrinsic polarisation fraction of $p$ = 0.93 $\pm$ 0.05 and $\mathrm{\sigma_{RM}}$ = 5.2 $\pm$ 0.3 rad m$^{-2}$.}
    \label{fig:251024depol}
\end{figure}

Here we report two additional sources within our newly added sample of 30 polarised non-repeating bursts that show evidence for spectral depolarisation, FRB 20250607C and FRB 20251024B, which brings the occurrence rate lower limit of spectral depolarisation in the CRAFT sample up to $\sim$6\%. FRB 20250607C is the subject of a separate study and will be described in detail by Balzan et al. (in prep). We model spectral depolarisation in FRB 20251024B using the modified Burns law:

\begin{equation}
\label{eq: burnslaw}
    p_{m}(\nu) = p \cdot \mathrm{exp\big(-2\sigma_{RM}^{2}\nu^{-4}\,c^{4}\big)},
\end{equation}

where $p_{m}(\nu)$ is the measured frequency dependent P/I fraction and $p$ is the intrinsic P/I fraction. To measure the polarisation fraction as a function of frequency, we first formed a one-dimensional spectrum by taking a boxcar average over the leading 4.4 ms of the FRB 20251024B pulse. We then averaged to a frequency resolution 32 MHz, and calculated P/I for each of these channels. Since we do not know the intrinsic polarisation properties of the burst, we set the priors on $p$ to between 0.0 and 1.0. We also provided a range of priors for $\mathrm{\sigma_{RM}}$ between 0.0 rad m$^{-2}$ and 100.0 rad m$^{-2}$. We obtained best fitting values of $p$ = 0.93 $\pm$ 0.05 and $\mathrm{\sigma_{RM}}$ = 5.2 $\pm$ 0.3 rad m$^{-2}$. As seen in Fig.~\ref{fig:251024depol}, there is clear evidence of depolarisation at lower frequency, to which the modified Burns law provides an acceptable fit (chi squared of 5.65 for 7 degrees of freedom), whereas a constant polarisation fraction is ruled out by the data (chi squared of 44.7 for 8 degrees of freedom).

With sufficiently strong turbulence, an incident FRB may be completely depolarised even at the upper edge of the ASKAP observing band. As discussed in Section~\ref{sec:unpol}, seven bursts fit this description. We investigated potential depolarisation by measuring a lower limit for $\mathrm{\sigma_{RM}}$ for each burst. To do so, we use Equation \ref{eq: burnslaw} and set $p$ to be unity and $p_{m}$ (at the maximum burst frequency, $\nu_{\mathrm{max}}$) to be the upper limit on the polarisation fraction. In Table \ref{tab:depol} we report the lower limits on $\sigma_{\mathrm{RM}}$ for each unpolarised burst in the \newhtr{} sample along with exact measurements for non-repeating bursts with confirmed spectral depolarisation. The typical unpolarised FRB in the sample requires a minimum $\mathrm{\sigma_{RM}}$ of $\sim$10 rad m$^{-2}$ to completely depolarise the burst. Reports on spectral depolarisation in repeating FRBs \citep{feng2022frequency, Mckinven_2023, kumar2023} show upper limits on spectral depolarisation typically $<$10 rad m$^{-2}$, although several sources such as FRB 20121102A \citep{hilmarsson2021rotation} and FRB 20190520B \citep{feng2022frequency} exhibit much stronger depolarisation of $\mathrm{\sigma_{RM}}$ = 30.9 rad m$^{-2}$ and 218.9 rad m$^{-2}$ respectively. An RM scattering origin for these depolarised FRBs would therefore suggest an unexpected population of faint and wide non-repeating bursts with high $\mathrm{\sigma_{RM}}$ values that have previously been associated with complex, dense, and highly magnetised circumburst environments. However, the current sample is only suggestive and further direct detections of $\sigma_{\mathrm{RM}}$ for more non-repeating FRBs are sorely needed.


One avenue for further insight into the origin of the frequency dependent polarisation seen here is the association (or lack thereof) of persistent radio sources to the depolarised samples. If the mechanism for generating both spectral depolarisation and a PRS is linked \citep{niu2022repeating, Yang_2024}, we should expect to see a PRS associated with some non-repeaters. This motivates continued deep searches for PRSs associated with depolarised non-repeating FRBs. If none are found, then the apparent $\mathrm{\sigma_{RM}}$ may be due to the intrinsic emission mechanism and not multi-path propagation. In this case, high S/N detections of non-repeating FRBs with wideband instruments may prove illustrative, given the potential to show deviations from the frequency dependence predicted by multipath propagation models.

\begin{table}
\centering
\setlength\tabcolsep{18pt}  
    \begin{threeparttable}
    \caption{Upper table: Lower limits on $\sigma_{\mathrm{RM}}$ for seven unpolarised bursts in the \newhtr{} sample. Lower table: Measurements of $\sigma_{\mathrm{RM}}$ for ASKAP bursts with confirmed spectral depolarisation.}
    \label{tab:depol}
    \begin{tabular}{ccc}
    \hline
    FRB     &  $\nu_{\mathrm{max}}$ (MHz)  &     $\mathrm{\sigma_{RM}}$ (rad m$^{-2}$) \\ \hline
    20240807A     & 1085.5 &  $>$9.3 \\
    20241226C     & 1071.5 &  $>$8.7 \\
    20250313A     & 1085.5 &  $>$11.5 \\
    20251010B     & 995.5 &   $>$8.6\\
    20251214A     & 948.5 &   $>$6.6 \\
    20260409E     & 1017.5 &  $>$7.6 \\ 
    20260515E     & 1439.5 &  $>$21.9 \\ \hline
    20230526A\tnote{a}   &   1439.5      &   12.6 $\pm$ 0.3 \\
    20250607C\tnote{b}   &   1081.5      &   3.56 $\pm$ 0.03 \\
    20251024B    &   1031.5      &   5.2 $\pm$ 0.3 \\ \hline
    \end{tabular}

      \begin{tablenotes}
      \small
      \item[a] \cite{2026MNRAS.545f1997U}.
      \item[b] Balzan et al., in prep.
    \end{tablenotes}
    \end{threeparttable}

\end{table}

\subsection{RM$_{\mathrm{EG}}$ Distribution}

\begin{figure*}
    \centering
    \subfigure[]{\includegraphics[width=0.49\linewidth]{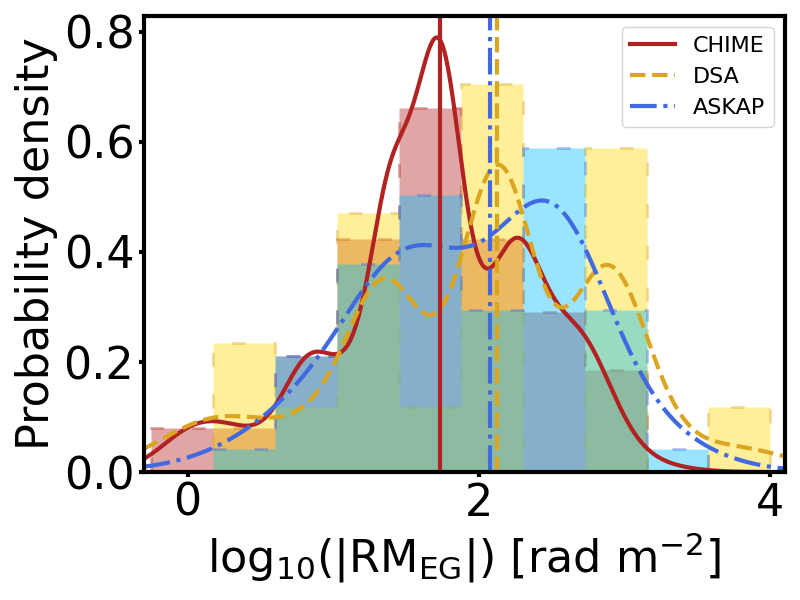}}
    \subfigure[]{\includegraphics[width=0.49\linewidth]{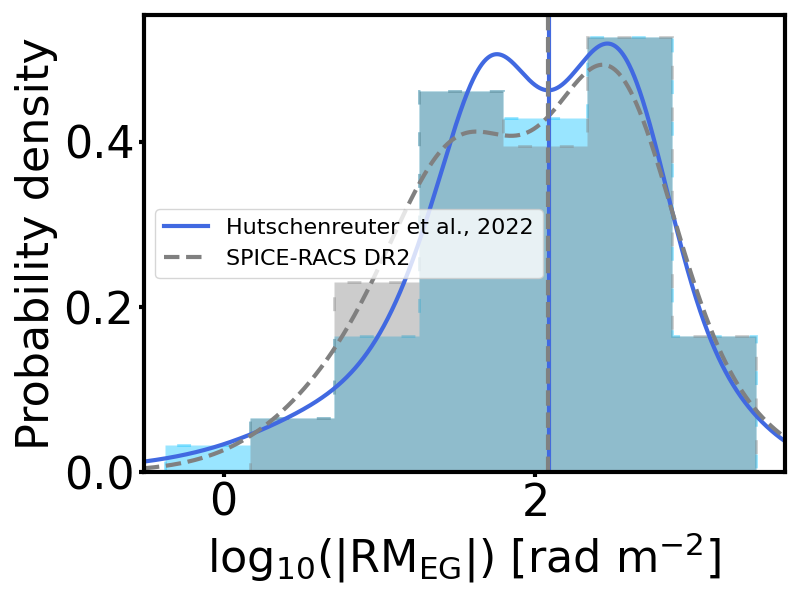}}
    \caption{Normalised histograms of $\mathrm{|RM_{EG}|}$ are shown as shaded regions. The bold curves and vertical lines show the KDE and median $\mathrm{|RM_{EG}|}$ respectively. (a) Absolute extra-galactic RM distribution for CHIME, DSA and ASKAP FRB samples. (b) Comparison between the ASKAP sample when corrected using the \protect\cite{hutschenreuter2022galactic} Galactic Faraday map and SPICE-RACS DR2 \protect\citep{thomson2026rapid}. Vertical lines show median RM.}
    \label{fig:rmhist}
\end{figure*}

We compare the histograms of the logarithm of the absolute extragalactic RM, $\log_{10}(|\rm RM_{EG}|)$, for ASKAP, CHIME and DSA in Fig.~\ref{fig:rmhist} (panel a). We also include density curves of the underlying distribution which were derived using a Kernel Density Estimator (KDE) on the log distributions as outlined in Appendix~\ref{sec:KDE}. We report both the geometric mean and median \rmeg\ for each sample in Table~\ref{tab:RMtable}, along with their uncertainties which were derived using a bootstrapping method also detailed in Appendix~\ref{sec:KDE}. As described in Appendix~\ref{sec:KDE}, comparable results are also obtained when using a straightforward calculation of the mean and standard deviation of the unweighted data points from each sample.
For the CHIME and DSA samples, RM$_{\mathrm{EG}}$ was corrected using the \cite{hutschenreuter2022galactic} Faraday map. \cite{sherman2023deep} only report RM$_{\mathrm{MW}}$ for a subset of the DSA bursts. Therefore, we calculate RM$_{\mathrm{MW}}$ for the remaining bursts using the \cite{hutschenreuter2022galactic} Faraday map; the SPICE-RACS DR2 Faraday map was not used for CHIME and DSA due to a declination limit of $+49^{\circ}$. 


As seen in Table~\ref{tab:RMtable} and Fig.~\ref{fig:rmhist}, the CHIME distribution skews significantly lower compared to ASKAP and DSA. If we compare the ASKAP sample (median \rmeg\ = 119.8 rad m$^{-2}$) to that of CHIME (median \rmeg\ = 54.1 rad m$^{-2}$) using a two-sample Anderson-Darling test, we obtain a test statistic and $p$-value of 2.06 and 0.046 respectively. Similarly, if we compare DSA (median \rmeg\ = 132.1 rad m$^{-2}$) to CHIME, we obtain a test statistic and $p$-value of 1.67 and 0.07. Finally, comparing ASKAP and DSA, we get values of -0.45 and 0.59 for the test statistic and $p$-value respectively. This is unsurprising given that ASKAP and DSA generally observe at similar frequencies, higher than those of CHIME. If the ASKAP and DSA bursts are combined into a single sample and compared with the CHIME sample, we obtain a test statistic and $p$-value of 2.96 and 0.02. 

This establishes that our best estimate of the intrinsic RM distribution of FRBs seen by CHIME differs significantly from that seen by ASKAP. Several possibilities can be put forward to account for this. First, the cause could be observational: intra-channel RM smearing is strongly dependent on the observing frequency ($\propto f^{-2}$) and the channel resolution. For typical ASKAP FRBs at a channel resolution of 1.0 MHz and a central frequency of 1.0 GHz, $|$RM$|$ can be measured up to $\sim$10$^{4}$ rad m$^{-2}$ with no appreciable loss in polarised flux. Furthermore, we can channelise the voltage data to higher frequency resolution if needed to search over larger $|$RM$|$ ranges (see Section~\ref{sec:RMfitting}). While it varies significantly over the CHIME band, at the central frequency of 600 MHz and with a channel resolution of 0.39 KHz, a loss of $>$50\% of the polarised flux occurs for RMs $\gtrsim$5000 rad m$^{-2}$\citep{pandhi2024polarization}. This RM limit, however, is considerably higher than those seen in the ASKAP non-repeating FRB sample. The difference in the ASKAP and CHIME RM distributions is already apparent at the level of hundreds of rad m$^{-2}$, where at most only a small fraction of polarised flux is lost for a CHIME FRB. 

A second possibility is that the difference is due to propagation effects: RM scattering in a turbulent plasma similarly scales with wavelength squared but also the magnitude of the RM incured through the plasma. This could lead to high RM bursts being completely depolarised before they reach the telescope. At face value, the results are consistent with this, with both ASKAP and DSA seeing a higher median \rmeg\ than CHIME. However, as shown in Fig.~\ref{fig:rm_v_li}, while a negative correlation is seen between \rmeg\ and L/I for both CHIME and ASKAP, it is not significant, with Spearman rank coefficients of -0.052 \citep{pandhi2024polarization} and -0.345 respectively.

To ensure that the observed differences between ASKAP and CHIME are not a result of the different Faraday maps used to correct for the galactic RM, we compare the ASKAP sample's \rmeg\ distribution derived using either the SPICE-RACS DR2 or \cite{hutschenreuter2022galactic} Faraday map. The corresponding histograms are shown in panel (c) of Fig. \ref{fig:rmhist}. We find very similar median \rmeg\ values of 119.8 rad m$^{-2}$ and 121.7 rad m$^{-2}$ between the two corrected samples. Therefore, we are confident that the differences in the \rmeg\ distribution between the various FRB samples are not artifacts of the choice of Faraday map used to account for RM$_{\mathrm{MW}}$.


\begin{table}
    \centering
    \setlength\extrarowheight{4pt}
    \setlength\tabcolsep{2pt}
    \caption{Arithmetic Median and geometric mean RMs for different non-repeating burst samples. Errors are reported as 68\% confidence intervals.}
    \begin{tabular}{p{2.5cm}ccc}
    \hline
        Sample & n & \thead{Geomtric mean RM$_{\mathrm{EG}}$\\(rad m$^{-2}$)} & \thead{Median RM$_{\mathrm{EG}}$\\(rad m$^{-2}$)} \\ \hline
        CHIME & 89 & 50.0$^{+10.1}_{-7.6}$  &   54.1$^{+5.6}_{-9.1}$ \\
        DSA   & 20 &   96.8$^{+50.7}_{-36.8}$ &   132.1$^{+36.9}_{-58.8}$ \\
        ASKAP  & 56 &   91.7$^{+21.4}_{-19.7}$  &   119.8$^{+32.4}_{-52.0}$ \\ [1ex]
        \hline 
        ASKAP \newline [Hutschenreuter] & 56 & 96.6$^{+24.0}_{-20.4}$ & 121.7$^{+30.8}_{-47.5}$ \\ [1ex]
        \hline
    \end{tabular}
    \label{tab:RMtable}
\end{table}


\section{Conclusion}
\label{sec:conclusion}

This work presents the polarimetry of the updated catalogue of 64 CRAFT HTR FRBs detected by ASKAP. We have shown that the fractional linear and circular polarisation distributions of the latest ASKAP sample are consistent with the CHIME and DSA sample. DSA bursts seem to possess slightly higher circular polarisation on average which requires further investigation. Furthermore, the value of \rmeg\ seen for CHIME FRBs is significantly lower on average than those for ASKAP or DSA FRBs, but neither intra-channel depolarisation nor RM scattering appears likely to be the reason. Of the new results from \newhtr{}, we show that a small but non-negligible fraction of non-repeating bursts show evidence for spectral depolarisation, and we show for the first time a potential dependence of polarisation fraction on width and/or spectral luminosity, with the widest ASKAP bursts always exhibiting low polarisation, and the most luminous bursts always exhibiting high polarisation.

More spectroscopic follow-up and host galaxy association will enable us to better constrain this possible correlation, as the sample size of CRACO-detected ASKAP FRBs increases. Fully characterising the dependence of FRB polarisation on burst width and luminosity will both provide substantial constraints on the emission mechanisms that drive FRBs, and provide a significant boost to the use of FRBs to probe cosmology and large-scale structure in the late-time Universe.

\section*{Acknowledgements}

M.G. is supported by the Australian Government through the Australian Research Council’s Discovery Projects funding scheme (DP210102103), and through UK STFC Grant ST/Y001117/1. M.G. acknowledges support from the Inter-University Institute for Data Intensive Astronomy (IDIA). IDIA is a partnership of the University of Cape Town, the University of Pretoria and the University of the Western Cape. For the purpose of open access, the author has applied a Creative Commons Attribution (CC BY) licence to any Author Accepted Manuscript version arising from this submission. 
A.B. acknowledges support through project CORTEX (NWA.1160.18.316) of the research programme NWA-ORC which is financed by the Dutch Research Council (NWO).
RMS, ATD and JJ-S acknowledges support through Australian Research Council Discovery Project DP220102305.
ZW acknowledges support from the Australian Government through the Australian Research Council Discovery Project DP250102020.
\section*{Data Availability}

The processed data products for each FRB are on the NT supercomputer and can be made available upon reasonable request.



\bibliographystyle{mnras}
\bibliography{bib} 

@article{hotan2021australian,
  title={Australian square kilometre array pathfinder: I. system description},
  author={Hotan, AW and Bunton, JD and Chippendale, AP and Whiting, M and Tuthill, John and Moss, Vanessa A and McConnell, D and Amy, SW and Huynh, MT and Allison, JR and others},
  journal={Publications of the Astronomical Society of Australia},
  volume={38},
  pages={e009},
  year={2021},
  publisher={Cambridge University Press}
}

@article{shannon2024commensal,
  title={The commensal real-time ASKAP fast transient incoherent-sum survey},
  author={Shannon, Ryan M and Bannister, Keith W and Bera, Apurba and Bhandari, Shivani and Day, Cherie K and Deller, Adam T and Dial, Tyson and Dobie, Dougal and Ekers, Ron D and Fong, Wen-fai and others},
  journal={Publications of the Astronomical Society of Australia},
  volume={42},
  pages={e036},
  year={2025},
  publisher={Cambridge University Press}
}

@article{scott2023celebi,
  title={CELEBI: The CRAFT Effortless Localisation and Enhanced Burst Inspection pipeline},
  author={Scott, Danica Rachel and Cho, Hyerin and Day, Cherie K and Deller, Adam T and Glowacki, Marcin and Gourdji, Kelly and Bannister, Keith W and Bera, Apurba and Bhandari, Shivani and James, Clancy W and others},
  journal={Astronomy and Computing},
  pages={100724},
  year={2023},
  publisher={Elsevier}
}

@article{dial2025frb,
  title={FRB 20230708A, a quasi-periodic FRB with unique temporal-polarimetric morphology},
  author={Dial, T and Deller, AT and Uttarkar, PA and Lower, ME and Shannon, RM and Gourdji, Kelly and Marnoch, Lachlan and Bera, A and Ryder, Stuart D and Glowacki, Marcin and others},
  journal={Monthly Notices of the Royal Astronomical Society},
  volume={536},
  number={4},
  pages={3220--3231},
  year={2025},
  publisher={Oxford University Press}
}

@article{sutinjo2023calculation,
  title={Calculation and Uncertainty of Fast Radio Burst Structure Based on Smoothed Data},
  author={Sutinjo, Adrian T and Scott, Danica R and James, Clancy W and Glowacki, Marcin and Bannister, Keith W and Cho, Hyerin and Day, Cherie K and Deller, Adam T and Perrett, Timothy P and Shannon, Ryan M},
  journal={The Astrophysical Journal},
  volume={954},
  number={1},
  pages={37},
  year={2023},
  publisher={IOP Publishing}
}

@article{anna2023magnetic,
  title={Magnetic field reversal in the turbulent environment around a repeating fast radio burst},
  author={Anna-Thomas, Reshma and Connor, Liam and Dai, Shi and Feng, Yi and Burke-Spolaor, Sarah and Beniamini, Paz and Yang, Yuan-Pei and Zhang, Yong-Kun and Aggarwal, Kshitij and Law, Casey J and others},
  journal={Science},
  volume={380},
  number={6645},
  pages={599--603},
  year={2023},
  publisher={American Association for the Advancement of Science}
}

@article{michilli2018extreme,
  title={An extreme magneto-ionic environment associated with the fast radio burst source FRB 121102},
  author={Michilli, Daniele and Seymour, A and Hessels, JWT and Spitler, LG and Gajjar, V and Archibald, AM and Bower, GC and Chatterjee, S and Cordes, JM and Gourdji, K and others},
  journal={Nature},
  volume={553},
  number={7687},
  pages={182--185},
  year={2018},
  publisher={Nature Publishing Group UK London}
}

@article{scott2025high,
  title={High-time-resolution properties of 35 fast radio bursts detected by the Commensal Real-time ASKAP Fast Transients survey},
  author={Scott, Danica R and Dial, Tyson and Bera, Apurba and Deller, Adam T and Glowacki, Marcin and Gourdji, Kelly and James, Clancy W and Shannon, Ryan M and Bannister, Keith W and Ekers, Ron D and others},
  journal={Publications of the Astronomical Society of Australia},
  volume={42},
  pages={e133},
  year={2025},
  publisher={Cambridge University Press}
}

@article{karastergiou2003v,
  title={| V|: new insight into the circular polarization of radio pulsars},
  author={Karastergiou, Aris and Johnston, S and Mitra, D and Van Leeuwen, AGJ and Edwards, RT},
  journal={Monthly Notices of the Royal Astronomical Society},
  volume={344},
  number={4},
  pages={L69--L73},
  year={2003},
  publisher={Blackwell Science Ltd Oxford, UK}
}

@article{oswald2023pulsar1,
  title={Pulsar polarization: a broad-band population view with the Parkes Ultra-Wideband receiver},
  author={Oswald, LS and Johnston, S and Karastergiou, A and Dai, Shi and Kerr, M and Lower, ME and Manchester, RN and Shannon, RM and Sobey, C and Weltevrede, P},
  journal={Monthly Notices of the Royal Astronomical Society},
  volume={520},
  number={4},
  pages={4961--4980},
  year={2023},
  publisher={Oxford University Press}
}

@article{brentjens2005faraday,
  title={Faraday rotation measure synthesis},
  author={Brentjens, Michiel A and De Bruyn, Antonius Gerardus},
  journal={Astronomy \& Astrophysics},
  volume={441},
  number={3},
  pages={1217--1228},
  year={2005},
  publisher={EDP Sciences}
}

@article{feng2022frequency,
  title={Frequency-dependent polarization of repeating fast radio bursts—implications for their origin},
  author={Feng, Yi and Li, Di and Yang, Yuan-Pei and Zhang, Yongkun and Zhu, Weiwei and Zhang, Bing and Lu, Wenbin and Wang, Pei and Dai, Shi and Lynch, Ryan S and others},
  journal={Science},
  volume={375},
  number={6586},
  pages={1266--1270},
  year={2022},
  publisher={American Association for the Advancement of Science}
}

@article{lu2023temporal,
  title={Temporal Evolution of Depolarization and Magnetic Field of Fast Radio Burst 20201124A},
  author={Lu, Wan-Jin and Zhao, Zhen-Yin and Wang, FY and Dai, ZG},
  journal={The Astrophysical Journal Letters},
  volume={956},
  number={1},
  pages={L9},
  year={2023},
  publisher={IOP Publishing}
}

@article{amiri2021first,
  title={The first CHIME/FRB fast radio burst catalog},
  author={Amiri, Mandana and Andersen, Bridget C and Bandura, Kevin and Berger, Sabrina and Bhardwaj, Mohit and Boyce, Michelle M and Boyle, PJ and Brar, Charanjot and Breitman, Daniela and Cassanelli, Tomas and others},
  journal={The Astrophysical Journal Supplement Series},
  volume={257},
  number={2},
  pages={59},
  year={2021},
  publisher={IOP Publishing}
}

@article{burn1966depolarization,
  title={On the depolarization of discrete radio sources by Faraday dispersion},
  author={Burn, BJ},
  journal={Monthly Notices of the Royal Astronomical Society},
  volume={133},
  number={1},
  pages={67--83},
  year={1966},
  publisher={Oxford University Press Oxford, UK}
}

@article{hilmarsson2021rotation,
  title={Rotation measure evolution of the repeating fast radio burst source FRB 121102},
  author={Hilmarsson, GH and Michilli, D and Spitler, LG and Wharton, RS and Demorest, P and Desvignes, G and Gourdji, K and Hackstein, S and Hessels, JWT and Nimmo, K and others},
  journal={The Astrophysical Journal Letters},
  volume={908},
  number={1},
  pages={L10},
  year={2021},
  publisher={IOP Publishing}
}

@article{li2026sudden,
  title={A sudden change and recovery in the magnetic environment around a repeating fast radio burst},
  author={Li, Y and Zhang, SB and Yang, YP and Tsai, CW and Yang, X and Law, CJ and Anna-Thomas, R and Chen, XL and Lee, KJ and Tang, ZF and others},
  journal={Science},
  volume={391},
  number={6782},
  pages={280--284},
  year={2026},
  publisher={American Association for the Advancement of Science}
}

@article{zhang2023fast,
  title={FAST observations of FRB 20220912A: burst properties and polarization characteristics},
  author={Zhang, Yong-Kun and Li, Di and Zhang, Bing and Cao, Shuo and Feng, Yi and Wang, Wei-Yang and Qu, Yuanhong and Niu, Jia-Rui and Zhu, Wei-Wei and Han, Jin-Lin and others},
  journal={The Astrophysical Journal},
  volume={955},
  number={2},
  pages={142},
  year={2023},
  publisher={IOP Publishing}
}

@article{zhang2025magnetar,
  title={The magnetar model's energy crisis for a prolific repeating fast radio burst source},
  author={Zhang, Jun-Shuo and Wang, Tian-Cong and Wang, Pei and Wu, Qin and Li, Di and Zhu, Weiwei and Zhang, Bing and Gao, He and Lee, Ke-Jia and Han, Jinlin and others},
  journal={arXiv preprint arXiv:2507.14707},
  year={2025}
}

@article{zhang2022fast,
  title={FAST observations of an extremely active episode of FRB 20201124A. II. Energy distribution},
  author={Zhang, Yong-Kun and Wang, Pei and Feng, Yi and Zhang, Bing and Li, Di and Tsai, Chao-Wei and Niu, Chen-Hui and Luo, Rui and Yao, Ju-Mei and Zhu, Wei-Wei and others},
  journal={Research in Astronomy and Astrophysics},
  volume={22},
  number={12},
  pages={124002},
  year={2022},
  publisher={IOP Publishing}
}

@article{macquart2020census,
  title={A census of baryons in the Universe from localized fast radio bursts},
  author={Macquart, J-P and Prochaska, JX and McQuinn, M and Bannister, KW and Bhandari, S and Day, CK and Deller, AT and Ekers, RD and James, CW and Marnoch, L and others},
  journal={Nature},
  volume={581},
  number={7809},
  pages={391--395},
  year={2020},
  publisher={Nature Publishing Group UK London}
}

@article{wang2025craft,
  title={The CRAFT coherent (CRACO) upgrade I: System description and results of the 110-ms radio transient pilot survey},
  author={Wang, Z and Bannister, KW and Gupta, V and Deng, X and Pilawa, M and Tuthill, J and Bunton, JD and Flynn, C and Glowacki, M and Jaini, A and others},
  journal={Publications of the Astronomical Society of Australia},
  volume={42},
  pages={e005},
  year={2025},
  publisher={Cambridge University Press}
}

@article{pandhi2024polarization,
  title={Polarization properties of 128 nonrepeating fast radio bursts from the first CHIME/FRB baseband catalog},
  author={Pandhi, Ayush and Pleunis, Ziggy and Mckinven, Ryan and Gaensler, BM and Su, Jianing and Ng, Cherry and Bhardwaj, Mohit and Brar, Charanjot and Cassanelli, Tomas and Cook, Amanda and others},
  journal={The Astrophysical Journal},
  volume={968},
  number={2},
  pages={50},
  year={2024},
  publisher={IOP Publishing}
}

@article{wu2025universal,
  title={A universal break in energy functions of three hyperactive repeating fast radio bursts},
  author={Wu, Q and Wang, FY and Zhao, ZY and Wang, P and Xu, H and Zhang, YK and Zhou, DJ and Niu, JR and Wang, WY and Yi, SX and others},
  journal={The Astrophysical Journal Letters},
  volume={979},
  number={2},
  pages={L42},
  year={2025},
  publisher={IOP Publishing}
}

@article{lorimer2007bright,
  title={A bright millisecond radio burst of extragalactic origin},
  author={Lorimer, Duncan R and Bailes, Matthew and McLaughlin, Maura Ann and Narkevic, David J and Crawford, Froney},
  journal={Science},
  volume={318},
  number={5851},
  pages={777--780},
  year={2007},
  publisher={American Association for the Advancement of Science}
}

@misc{glowacki2026pinkupdateimprovementscelebi,
      title={A PINK update: Improvements to the CELEBI fast radio burst data reduction and analysis pipeline}, 
      author={M. Glowacki and T. Dial and A. Bera and A. T. Deller and K. Gourdji and A. Jaini and D. Scott and Y. Wang and K. Desnos and A. C. Gordon and R. L. Davies and R. M. Shannon},
      year={2026},
      eprint={2605.06766},
      archivePrefix={arXiv},
      primaryClass={astro-ph.IM},
      url={https://arxiv.org/abs/2605.06766}, 
}

@article{hutschenreuter2022galactic,
  title={The galactic Faraday rotation sky 2020},
  author={Hutschenreuter, Sebastian and Anderson, Craig S and Betti, Sarah and Bower, Geoffrey C and Brown, J-A and Br{\"u}ggen, Marcus and Carretti, Ettore and Clarke, Tracy and Clegg, Andrew and Costa, Allison and others},
  journal={Astronomy \& Astrophysics},
  volume={657},
  pages={A43},
  year={2022},
  publisher={EDP Sciences}
}

@article{day2020high,
  title={High time resolution and polarization properties of ASKAP-localized fast radio bursts},
  author={Day, Cherie K and Deller, Adam T and Shannon, Ryan M and Qiu, Hao and Bannister, Keith W and Bhandari, Shivani and Ekers, Ron and Flynn, Chris and James, Clancy W and Macquart, Jean-Pierre and others},
  journal={Monthly Notices of the Royal Astronomical Society},
  volume={497},
  number={3},
  pages={3335--3350},
  year={2020},
  publisher={Oxford University Press}
}

@article{mckinnon2025polarization,
  title={Polarization Estimation for Radio Pulsars},
  author={McKinnon, MM},
  journal={The Astrophysical Journal},
  volume={982},
  number={2},
  pages={136},
  year={2025},
  publisher={The American Astronomical Society}
}

@article{kirsten2024link,
  title={A link between repeating and non-repeating fast radio bursts through their energy distributions},
  author={Kirsten, F and Ould-Boukattine, OS and Herrmann, W and Gawro{\'n}ski, MP and Hessels, JWT and Lu, W and Snelders, MP and Chawla, P and Yang, J and Blaauw, R and others},
  journal={Nature Astronomy},
  volume={8},
  number={3},
  pages={337--346},
  year={2024},
  publisher={Nature Publishing Group UK London}
}

@article{niu2022repeating,
  title={A repeating fast radio burst associated with a persistent radio source},
  author={Niu, C-H and Aggarwal, K and Li, D and Zhang, X and Chatterjee, S and Tsai, C-W and Yu, W and Law, CJ and Burke-Spolaor, S and Cordes, JM and others},
  journal={Nature},
  volume={606},
  number={7916},
  pages={873--877},
  year={2022},
  publisher={Nature Publishing Group UK London}
}

@article{bruni2024nebular,
  title={A nebular origin for the persistent radio emission of fast radio bursts},
  author={Bruni, Gabriele and Piro, Luigi and Yang, Yuan-Pei and Quai, Salvatore and Zhang, Bing and Palazzi, Eliana and Nicastro, Luciano and Feruglio, Chiara and Tripodi, Roberta and O’Connor, Brendan and others},
  journal={Nature},
  volume={632},
  number={8027},
  pages={1014--1016},
  year={2024},
  publisher={Nature Publishing Group UK London}
}

@article{bruni2025discovery,
  title={Discovery of a persistent radio source associated with FRB 20240114A},
  author={Bruni, G and Piro, L and Yang, Y-P and Palazzi, E and Nicastro, L and Rossi, A and Savaglio, S and Maiorano, E and Zhang, B},
  journal={Astronomy \& Astrophysics},
  volume={695},
  pages={L12},
  year={2025},
  publisher={EDP Sciences}
}

@article{sherman2024deep,
  title={Deep Synoptic Array Science: Polarimetry of 25 New Fast Radio Bursts Provides Insights into Their Origins},
  author={Sherman, Myles B and Connor, Liam and Ravi, Vikram and Law, Casey and Chen, Ge and Catha, Morgan and Faber, Jakob T and Hallinan, Gregg and Harnach, Charlie and Hellbourg, Greg and others},
  journal={The Astrophysical Journal},
  volume={964},
  number={2},
  pages={131},
  year={2024},
  publisher={The American Astronomical Society}
}

@misc{dial2026,
      title={FRB20250613A: a remarkable repeating FRB with apparent millisecond-timescale scattering variations}, 
      author={T. Dial and A. T. Deller and Alexa C. Gordon and P. A. Uttarkar and R. M. Shannon and Ziteng Wang and M. Caleb and Wen-fai Fong and Marcin Glowacki and Kelly Gourdji and Joscha N. Jahns-Schindler},
      year={2026},
      eprint={2607.00505},
      archivePrefix={arXiv},
      primaryClass={astro-ph.HE},
      url={https://arxiv.org/abs/2607.00505}, 
}

@article{cho2020spectropolarimetric,
  title={Spectropolarimetric analysis of FRB 181112 at microsecond resolution: implications for fast radio burst emission mechanism},
  author={Cho, Hyerin and Macquart, Jean-Pierre and Shannon, Ryan M and Deller, Adam T and Morrison, Ian S and Ekers, Ron D and Bannister, Keith W and Farah, Wael and Qiu, Hao and Sammons, Mawson W and others},
  journal={The Astrophysical Journal Letters},
  volume={891},
  number={2},
  pages={L38},
  year={2020},
  publisher={The American Astronomical Society}
}

@ARTICLE{aghanim2020planck,
       author = {{Planck Collaboration} and {Aghanim}, N. and {Akrami}, Y. and {Ashdown}, M. and {Aumont}, J. and {Baccigalupi}, C. and {Ballardini}, M. and {Banday}, A.~J. and {Barreiro}, R.~B. and {Bartolo}, N. and {Basak}, S. and {Battye}, R. and {Benabed}, K. and {Bernard}, J.-P. and {Bersanelli}, M. and {Bielewicz}, P. and {Bock}, J.~J. and {Bond}, J.~R. and {Borrill}, J. and {Bouchet}, F.~R. and {Boulanger}, F. and {Bucher}, M. and {Burigana}, C. and {Butler}, R.~C. and {Calabrese}, E. and {Cardoso}, J.-F. and {Carron}, J. and {Challinor}, A. and {Chiang}, H.~C. and {Chluba}, J. and {Colombo}, L.~P.~L. and {Combet}, C. and {Contreras}, D. and {Crill}, B.~P. and {Cuttaia}, F. and {de Bernardis}, P. and {de Zotti}, G. and {Delabrouille}, J. and {Delouis}, J.-M. and {Di Valentino}, E. and {Diego}, J.~M. and {Dor{\'e}}, O. and {Douspis}, M. and {Ducout}, A. and {Dupac}, X. and {Dusini}, S. and {Efstathiou}, G. and {Elsner}, F. and {En{\ss}lin}, T.~A. and {Eriksen}, H.~K. and {Fantaye}, Y. and {Farhang}, M. and {Fergusson}, J. and {Fernandez-Cobos}, R. and {Finelli}, F. and {Forastieri}, F. and {Frailis}, M. and {Fraisse}, A.~A. and {Franceschi}, E. and {Frolov}, A. and {Galeotta}, S. and {Galli}, S. and {Ganga}, K. and {G{\'e}nova-Santos}, R.~T. and {Gerbino}, M. and {Ghosh}, T. and {Gonz{\'a}lez-Nuevo}, J. and {G{\'o}rski}, K.~M. and {Gratton}, S. and {Gruppuso}, A. and {Gudmundsson}, J.~E. and {Hamann}, J. and {Handley}, W. and {Hansen}, F.~K. and {Herranz}, D. and {Hildebrandt}, S.~R. and {Hivon}, E. and {Huang}, Z. and {Jaffe}, A.~H. and {Jones}, W.~C. and {Karakci}, A. and {Keih{\"a}nen}, E. and {Keskitalo}, R. and {Kiiveri}, K. and {Kim}, J. and {Kisner}, T.~S. and {Knox}, L. and {Krachmalnicoff}, N. and {Kunz}, M. and {Kurki-Suonio}, H. and {Lagache}, G. and {Lamarre}, J.-M. and {Lasenby}, A. and {Lattanzi}, M. and {Lawrence}, C.~R. and {Le Jeune}, M. and {Lemos}, P. and {Lesgourgues}, J. and {Levrier}, F. and {Lewis}, A. and {Liguori}, M. and {Lilje}, P.~B. and {Lilley}, M. and {Lindholm}, V. and {L{\'o}pez-Caniego}, M. and {Lubin}, P.~M. and {Ma}, Y.-Z. and {Mac{\'\i}as-P{\'e}rez}, J.~F. and {Maggio}, G. and {Maino}, D. and {Mandolesi}, N. and {Mangilli}, A. and {Marcos-Caballero}, A. and {Maris}, M. and {Martin}, P.~G. and {Martinelli}, M. and {Mart{\'\i}nez-Gonz{\'a}lez}, E. and {Matarrese}, S. and {Mauri}, N. and {McEwen}, J.~D. and {Meinhold}, P.~R. and {Melchiorri}, A. and {Mennella}, A. and {Migliaccio}, M. and {Millea}, M. and {Mitra}, S. and {Miville-Desch{\^e}nes}, M.-A. and {Molinari}, D. and {Montier}, L. and {Morgante}, G. and {Moss}, A. and {Natoli}, P. and {N{\o}rgaard-Nielsen}, H.~U. and {Pagano}, L. and {Paoletti}, D. and {Partridge}, B. and {Patanchon}, G. and {Peiris}, H.~V. and {Perrotta}, F. and {Pettorino}, V. and {Piacentini}, F. and {Polastri}, L. and {Polenta}, G. and {Puget}, J.-L. and {Rachen}, J.~P. and {Reinecke}, M. and {Remazeilles}, M. and {Renzi}, A. and {Rocha}, G. and {Rosset}, C. and {Roudier}, G. and {Rubi{\~n}o-Mart{\'\i}n}, J.~A. and {Ruiz-Granados}, B. and {Salvati}, L. and {Sandri}, M. and {Savelainen}, M. and {Scott}, D. and {Shellard}, E.~P.~S. and {Sirignano}, C. and {Sirri}, G. and {Spencer}, L.~D. and {Sunyaev}, R. and {Suur-Uski}, A.-S. and {Tauber}, J.~A. and {Tavagnacco}, D. and {Tenti}, M. and {Toffolatti}, L. and {Tomasi}, M. and {Trombetti}, T. and {Valenziano}, L. and {Valiviita}, J. and {Van Tent}, B. and {Vibert}, L. and {Vielva}, P. and {Villa}, F. and {Vittorio}, N. and {Wandelt}, B.~D. and {Wehus}, I.~K. and {White}, M. and {White}, S.~D.~M. and {Zacchei}, A. and {Zonca}, A.},
        title = "{Planck 2018 results. VI. Cosmological parameters}",
      journal = {\aap},
         year = 2020,
        month = sep,
       volume = {641},
          eid = {A6},
        pages = {A6},
          doi = {10.1051/0004-6361/201833910},
archivePrefix = {arXiv},
       eprint = {1807.06209},
 primaryClass = {astro-ph.CO},
       adsurl = {https://ui.adsabs.harvard.edu/abs/2020A&A...641A...6P}
}

@article{thomson2026rapid,
  title={The Rapid ASKAP Continuum Survey VII: Spectra and Polarisation In Cutouts of Extragalactic Sources (SPICE-RACS) Second Data Release--Unveiling the Magnetised Sky},
  author={Thomson, Alec JM and Galvin, Timothy J and Duchesne, Stefan W and Lenc, Emil and Heald, George and Hlinka, Ondrej and Malik, Sunil and Anderson, Craig S and Osinga, Erik and Baidoo, Lerato and others},
  journal={Publications of the Astronomical Society of Australia},
  pages={1--37},
  year={2026},
  publisher={Cambridge University Press}
}

@book{silverman2018density,
  title={Density estimation for statistics and data analysis},
  author={Silverman, Bernard W},
  year={2018},
  publisher={Routledge}
}

@ARTICLE{2026MNRAS.545f1997U,
       author = {{Uttarkar}, Pavan A. and {Shannon}, Ryan M. and {Gourdji}, Kelly and {Deller}, Adam T. and {Dial}, Tyson and {Glowacki}, Marcin and {Bera}, Apurba and {Gordon}, Alexa C. and {Ryder}, Stuart D. and {Tejos}, Nicolas and {Bhandari}, Shivani and {Wang}, Yuanming},
        title = "{A depolarization census of ASKAP fast radio bursts}",
      journal = {\mnras},
         year = 2026,
        month = jan,
       volume = {545},
       number = {2},
          eid = {staf1997},
        pages = {staf1997},
          doi = {10.1093/mnras/staf1997},
archivePrefix = {arXiv},
       eprint = {2503.19749},
 primaryClass = {astro-ph.HE},
       adsurl = {https://ui.adsabs.harvard.edu/abs/2026MNRAS.545f1997U}
}

@article{sherman2023deep,
  title={Deep Synoptic Array Science: Implications of Faraday Rotation Measures of Fast Radio Bursts Localized to Host Galaxies},
  author={Sherman, Myles B and Connor, Liam and Ravi, Vikram and Law, Casey and Chen, Ge and Sharma, Kritti and Catha, Morgan and Faber, Jakob T and Hallinan, Gregg and Harnach, Charlie and others},
  journal={The Astrophysical Journal Letters},
  volume={957},
  number={1},
  pages={L8},
  year={2023},
  publisher={The American Astronomical Society}
}

@article{ocker2026ne2025,
  title={NE2025: An Updated Electron Density Model for the Galactic Interstellar Medium},
  author={Ocker, Stella Koch and Cordes, James M},
  journal={The Astrophysical Journal},
  volume={1002},
  number={1},
  pages={3},
  year={2026},
  publisher={The American Astronomical Society}
}

@article{prochaska2019probing,
  title={Probing Galactic haloes with fast radio bursts},
  author={Prochaska, J Xavier and Zheng, Yong},
  journal={Monthly Notices of the Royal Astronomical Society},
  volume={485},
  number={1},
  pages={648--665},
  year={2019},
  publisher={Oxford University Press}
}

@article{keating2020exploring,
  title={Exploring the dispersion measure of the Milky Way halo},
  author={Keating, Laura C and Pen, Ue-Li},
  journal={Monthly Notices of the Royal Astronomical Society: Letters},
  volume={496},
  number={1},
  pages={L106--L110},
  year={2020},
  publisher={Oxford University Press}
}

@article{petroff2022fast,
  title={Fast radio bursts at the dawn of the 2020s},
  author={Petroff, E and Hessels, JWT and Lorimer, DR},
  journal={The Astronomy and Astrophysics Review},
  volume={30},
  number={1},
  pages={2},
  year={2022},
  publisher={Springer}
}

@article{melrose1971degree,
  title={On the degree of circular polarization of synchrotron radiation},
  author={Melrose, DB},
  journal={Astrophysics and Space Science},
  volume={12},
  number={1},
  pages={172--192},
  year={1971},
  publisher={Springer}
}

@article{jiang2025ninety,
  title={Ninety percent circular polarization detected in a repeating fast radio burst},
  author={Jiang, Jinchen and Xu, Jiangwei and Niu, Jiarui and Lee, Kejia and Zhu, Weiwei and Zhang, Bing and Qu, Yuanhong and Xu, Heng and Zhou, Dejiang and Cao, Shunshun and others},
  journal={National Science Review},
  volume={12},
  number={2},
  pages={nwae293},
  year={2025},
  publisher={Oxford University Press}
}

@article{beniamini2022faraday,
  title={Faraday depolarization and induced circular polarization by multipath propagation with application to FRBs},
  author={Beniamini, Paz and Kumar, Pawan and Narayan, Ramesh},
  journal={Monthly Notices of the Royal Astronomical Society},
  volume={510},
  number={3},
  pages={4654--4668},
  year={2022},
  publisher={Oxford University Press}
}

@book{condon2016essential,
  title={Essential radio astronomy},
  author={Condon, James Justin and Ransom, Scott M and Condon, James Justin and Condon, James J},
  series    = {Princeton Series in Modern Observational Astronomy},
  volume={2},
  year={2016},
  publisher={Princeton University Press Princeton, NJ}
}

@article{kaplan1958nonparametric,
  title={Nonparametric estimation from incomplete observations},
  author={Kaplan, Edward L and Meier, Paul},
  journal={Journal of the American statistical association},
  volume={53},
  number={282},
  pages={457--481},
  year={1958},
  publisher={Taylor \& Francis}
}

@article{Davidson-Pilon2019,
  doi = {10.21105/joss.01317},
  url = {https://doi.org/10.21105/joss.01317},
  year = {2019},
  publisher = {The Open Journal},
  volume = {4},
  number = {40},
  pages = {1317},
  author = {Cameron Davidson-Pilon},
  title = {lifelines: survival analysis in Python},
  journal = {Journal of Open Source Software}
}

@article{peto1972asymptotically,
  title={Asymptotically efficient rank invariant test procedures},
  author={Peto, Richard and Peto, Julian},
  journal={Journal of the Royal Statistical Society: Series A (General)},
  volume={135},
  number={2},
  pages={185--198},
  year={1972},
  publisher={Wiley Online Library}
}

@ARTICLE{chime2catalog,
       author = {{Chime/Frb Collaboration} and {Abbott}, Thomas and {Andersen}, Bridget C. and {Andrew}, Shion and {Bandura}, Kevin and {Bhardwaj}, Mohit and {Bhusare}, Yash and {Brar}, Charanjot and {Cassanelli}, Tomas and {Chatterjee}, Shami and {Cliche}, Jean-Francois and {Cook}, Amanda M. and {Curtin}, Alice and {Dobbs}, Matt and {Dong}, Fengqiu Adam and {Eadie}, Gwendolyn and {Eftekhari}, Tarraneh and {Fonseca}, Emmanuel and {Gaensler}, B.~M. and {Good}, Deborah and {Halpern}, Mark and {Hessels}, Jason W.~T. and {Ibik}, Adaeze and {Jain}, Naman and {Joseph}, Ronniy C. and {Kader}, Zarif and {Kaspi}, Victoria M. and {Khan}, Afrokk and {Kharel}, Bikash and {Kumar}, Ajay and {Landecker}, T.~L. and {Lang}, Dustin and {Lanman}, Adam E. and {L'Argent}, Magnus and {Lazda}, Mattias and {Leung}, Calvin and {Li}, Dong Zi and {Lintott}, Chris J. and {Main}, Robert and {Masui}, Kiyoshi W. and {Mate}, Sujay and {McGregor}, Kyle and {McKinven}, Ryan and {Mena-Parra}, Juan and {Meyers}, Bradley W. and {Michilli}, Daniele and {Ng}, Cherry and {Ng}, Mason and {Nimmo}, Kenzie and {Noble}, Gavin and {Pandhi}, Ayush and {Patil}, Swarali S. and {Pearlman}, Aaron B. and {Pen}, Ue-Li and {Pleunis}, Ziggy and {Prochaska}, J. Xavier and {Rafiei-Ravandi}, Masoud and {Ransom}, Scott and {Renard}, Andre and {Sammons}, Mawson W. and {Sand}, Ketan R. and {Scholz}, Paul and {Shah}, Vishwangi and {Shin}, Kaitlyn and {Siegel}, Seth R. and {Sirota}, Sloane and {Smith}, Kendrick and {Stairs}, Ingrid and {Stenning}, David C. and {Tendulkar}, Shriharsh P. and {Vanderlinde}, Keith and {Walmsley}, Mike and {Wang}, Haochen and {Wulf}, Dallas},
        title = "{The Second CHIME/FRB Catalog of Fast Radio Bursts}",
      journal = {\apjs},
         year = 2026,
        month = mar,
       volume = {283},
       number = {1},
          eid = {34},
        pages = {34},
          doi = {10.3847/1538-4365/ae3828},
archivePrefix = {arXiv},
       eprint = {2601.09399},
 primaryClass = {astro-ph.HE},
       adsurl = {https://ui.adsabs.harvard.edu/abs/2026ApJS..283...34C}
}

@article{Wang_2026,
doi = {10.1088/1674-4527/ae8428},
url = {https://doi.org/10.1088/1674-4527/ae8428},
year = {2026},
month = {jul},
publisher = {National Astromonical Observatories, CAS and IOP Publishing},
volume = {26},
number = {8},
pages = {084010},
author = {Wang, Fayin and Jia, Xuandong and Gao, Daohong and Dai, Zigao},
title = {Fast Radio Burst Cosmology: Hubble Tension and Dark Energy},
journal = {Research in Astronomy and Astrophysics}
}

@ARTICLE{CHIMEdesc,
       author = {{CHIME/FRB Collaboration} and {Amiri}, M. and {Bandura}, K. and {Berger}, P. and {Bhardwaj}, M. and {Boyce}, M.~M. and {Boyle}, P.~J. and {Brar}, C. and {Burhanpurkar}, M. and {Chawla}, P. and {Chowdhury}, J. and {Cliche}, J.-F. and {Cranmer}, M.~D. and {Cubranic}, D. and {Deng}, M. and {Denman}, N. and {Dobbs}, M. and {Fandino}, M. and {Fonseca}, E. and {Gaensler}, B.~M. and {Giri}, U. and {Gilbert}, A.~J. and {Good}, D.~C. and {Guliani}, S. and {Halpern}, M. and {Hinshaw}, G. and {H{\"o}fer}, C. and {Josephy}, A. and {Kaspi}, V.~M. and {Landecker}, T.~L. and {Lang}, D. and {Liao}, H. and {Masui}, K.~W. and {Mena-Parra}, J. and {Naidu}, A. and {Newburgh}, L.~B. and {Ng}, C. and {Patel}, C. and {Pen}, U.-L. and {Pinsonneault-Marotte}, T. and {Pleunis}, Z. and {Rafiei Ravandi}, M. and {Ransom}, S.~M. and {Renard}, A. and {Scholz}, P. and {Sigurdson}, K. and {Siegel}, S.~R. and {Smith}, K.~M. and {Stairs}, I.~H. and {Tendulkar}, S.~P. and {Vanderlinde}, K. and {Wiebe}, D.~V.},
        title = "{The CHIME Fast Radio Burst Project: System Overview}",
      journal = {\apj},
         year = 2018,
        month = aug,
       volume = {863},
       number = {1},
          eid = {48},
        pages = {48},
          doi = {10.3847/1538-4357/aad188},
archivePrefix = {arXiv},
       eprint = {1803.11235},
 primaryClass = {astro-ph.IM},
       adsurl = {https://ui.adsabs.harvard.edu/abs/2018ApJ...863...48C}
}

@article{van2026rm,
  title={RM-Tools: Software for Analyzing Polarized Radio Spectra},
  author={Van Eck, Cameron L and R. Purcell, Cormac and Baidoo, Lerato and Thomson, Alec JM and Ma, Yik Ki and Oberhelman, Lindsey and Osinga, Erik and Vanderwoude, Shannon and West, Jennifer L and Ideguchi, Shinsuke and others},
  journal={The Astrophysical Journal Supplement Series},
  volume={283},
  number={1},
  pages={28},
  year={2026},
  publisher={The American Astronomical Society}
}

@article{fine2023correcting,
  title={Correcting bandwidth depolarization by extreme Faraday rotation},
  author={Fine, Maxwell A and Van Eck, Cameron L and Pratley, Luke},
  journal={Monthly Notices of the Royal Astronomical Society},
  volume={520},
  number={4},
  pages={4822--4835},
  year={2023},
  publisher={Oxford University Press}
}

@ARTICLE{caleb2025,
       author = {{Caleb}, Manisha and {Nanayakkara}, Themiya and {Stappers}, Benjamin and {Pastor-Marazuela}, In{\'e}s and {Khrykin}, Ilya S. and {Glazebrook}, Karl and {Tejos}, Nicolas and {Prochaska}, J. Xavier and {Rajwade}, Kaustubh and {Mas-Ribas}, Lluis and {Driessen}, Laura N. and {Fong}, Wen-fai and {Gordon}, Alexa C. and {Hoffmann}, Jordan and {James}, Clancy W. and {Jankowski}, Fabian and {Kahinga}, Lordrick and {Kramer}, Michael and {Simha}, Sunil and {Barr}, Ewan D. and {Christiaan Bezuidenhout}, Mechiel and {Deng}, Xihan and {Lin}, Zeren and {Marnoch}, Lachlan and {Martin}, Christopher D. and {Nugent}, Anya and {Shaji}, Kavya and {Tian}, Jun},
        title = "{A fast radio burst from the first 3 billion years of the Universe}",
      journal = {arXiv e-prints},
         year = 2025,
        month = aug,
          eid = {arXiv:2508.01648},
        pages = {arXiv:2508.01648},
          doi = {10.48550/arXiv.2508.01648},
archivePrefix = {arXiv},
       eprint = {2508.01648},
 primaryClass = {astro-ph.HE},
       adsurl = {https://ui.adsabs.harvard.edu/abs/2025arXiv250801648C}
}

@article{baptista2024measuring,
  title={Measuring the Variance of the Macquart Relation in Redshift--Extragalactic Dispersion Measure Modeling},
  author={Baptista, Jay and Prochaska, J Xavier and Mannings, Alexandra G and James, CW and Shannon, RM and Ryder, Stuart D and Deller, AT and Scott, Danica R and Glowacki, Marcin and Tejos, Nicolas},
  journal={The Astrophysical Journal},
  volume={965},
  number={1},
  pages={57},
  year={2024},
  publisher={IOP Publishing}
}

@ARTICLE{sharma2026,
       author = {{Sharma}, Kritti and {Krause}, Elisabeth and {Ravi}, Vikram and {Reischke}, Robert and {Connor}, Liam and {R.~S.}, Pranjal and {Anbajagane}, Dhayaa},
        title = "{Probing Baryonic Feedback and Cosmology with the 3 {\texttimes} 2-point Statistic of Fast Radio Bursts and Galaxies}",
      journal = {\apj},
         year = 2026,
        month = feb,
       volume = {998},
       number = {1},
          eid = {109},
        pages = {109},
          doi = {10.3847/1538-4357/ae2ff9},
archivePrefix = {arXiv},
       eprint = {2509.05866},
 primaryClass = {astro-ph.CO},
       adsurl = {https://ui.adsabs.harvard.edu/abs/2026ApJ...998..109S}
}

@ARTICLE{kumar2022,
       author = {{Kumar}, P. and {Shannon}, R.~M. and {Lower}, M.~E. and {Bhandari}, S. and {Deller}, A.~T. and {Flynn}, C. and {Keane}, E.~F.},
        title = "{Circularly polarized radio emission from the repeating fast radio burst source FRB 20201124A}",
      journal = {\mnras},
         year = 2022,
        month = may,
       volume = {512},
       number = {3},
        pages = {3400-3413},
          doi = {10.1093/mnras/stac683},
archivePrefix = {arXiv},
       eprint = {2109.11535},
 primaryClass = {astro-ph.HE},
       adsurl = {https://ui.adsabs.harvard.edu/abs/2022MNRAS.512.3400K}
}

@ARTICLE{uttarkar2024,
       author = {{Uttarkar}, Pavan A. and {Shannon}, R.~M. and {Gourdji}, K. and {Deller}, A.~T. and {Day}, C.~K. and {Bhandari}, S.},
        title = "{Searching for the spectral depolarization of ASKAP one-off FRB sources}",
      journal = {\mnras},
         year = 2024,
        month = jan,
       volume = {527},
       number = {2},
        pages = {4285-4296},
          doi = {10.1093/mnras/stad3437},
archivePrefix = {arXiv},
       eprint = {2308.14387},
 primaryClass = {astro-ph.HE},
       adsurl = {https://ui.adsabs.harvard.edu/abs/2024MNRAS.527.4285U}
}

@article{Yang_2024,
doi = {10.3847/1538-4357/ad7d02},
url = {https://doi.org/10.3847/1538-4357/ad7d02},
year = {2024},
month = {nov},
publisher = {The American Astronomical Society},
volume = {976},
number = {2},
pages = {165},
author = {Yang, Ai Yuan and Feng, Yi and Tsai, Chao-Wei and Li, Di and Shi, Hui and Wang, Pei and Yang, Yuan-Pei and Zhang, Yong-Kun and Niu, Chen-Hui and Yao, Ju-Mei and Cui, Yu-Zhu and Su, Ren-Zhi and Li, Xiao-Feng and Zhang, Jun-Shuo and Zhu, Yu-Hao and Cotton, W. D.},
title = {The Variability of Persistent Radio Sources of Fast Radio Bursts},
journal = {The Astrophysical Journal}
}

@ARTICLE{james2025esd,
       author = {{James}, C.~W. and {Deller}, A.~T. and {Dial}, T. and {Glowacki}, M. and {Tingay}, S.~J. and {Bannister}, K.~W. and {Bera}, A. and {Bhat}, N.~D.~R. and {Ekers}, R.~D. and {Gupta}, V. and {Jaini}, A. and {Morgan}, J. and {Jahns-Schindler}, J.~N. and {Shannon}, R.~M. and {Sukhov}, M. and {Tuthill}, J. and {Wang}, Z.},
        title = "{A Nanosecond-duration Radio Pulse Originating from the Defunct Relay 2 Satellite}",
      journal = {\apjl},
         year = 2025,
        month = jul,
       volume = {987},
       number = {1},
          eid = {L16},
        pages = {L16},
          doi = {10.3847/2041-8213/ade3d3},
archivePrefix = {arXiv},
       eprint = {2506.11462},
 primaryClass = {astro-ph.EP},
       adsurl = {https://ui.adsabs.harvard.edu/abs/2025ApJ...987L..16J}
}

@ARTICLE{hoffman2026dm,
       author = {{Hoffmann}, Jordan Luke and {James}, Clancy and {Prochaska}, Xavier and {Glowacki}, Marcin},
        title = "{I can see your halo: Constraining the MilkyWay halo DM with FRB population studies}",
      journal = {\pasa},
         year = 2026,
        month = jan,
       volume = {43},
          eid = {e017},
        pages = {e017},
          doi = {10.1017/pasa.2026.10143},
archivePrefix = {arXiv},
       eprint = {2601.05496},
 primaryClass = {astro-ph.GA},
       adsurl = {https://ui.adsabs.harvard.edu/abs/2026PASA...43...17H}
}

@article{Cook_2023,
doi = {10.3847/1538-4357/acbbd0},
url = {https://doi.org/10.3847/1538-4357/acbbd0},
year = {2023},
month = {mar},
publisher = {The American Astronomical Society},
volume = {946},
number = {2},
pages = {58},
author = {Cook, Amanda M. and Bhardwaj, Mohit and Gaensler, B. M. and Scholz, Paul and Eadie, Gwendolyn M. and Hill, Alex S. and Kaspi, Victoria M. and Masui, Kiyoshi W. and Curtin, Alice P. and Dong, Fengqiu Adam and Fonseca, Emmanuel and Herrera-Martin, Antonio and Kaczmarek, Jane and Lanman, Adam E. and Lazda, Mattias and Leung, Calvin and Meyers, Bradley W. and Michilli, Daniele and Pandhi, Ayush and Pearlman, Aaron B. and Pleunis, Ziggy and Ransom, Scott and Rahman, Mubdi and Sand, Ketan R. and Shin, Kaitlyn and Smith, Kendrick and Stairs, Ingrid and Stenning, David C.},
title = {An FRB Sent Me a DM: Constraining the Electron Column of the Milky Way Halo with Fast Radio Burst Dispersion Measures from CHIME/FRB},
journal = {The Astrophysical Journal}
}

@article{Mckinven_2023,
doi = {10.3847/1538-4357/acd188},
url = {https://doi.org/10.3847/1538-4357/acd188},
year = {2023},
month = {jul},
publisher = {The American Astronomical Society},
volume = {951},
number = {1},
pages = {82},
author = {Mckinven, R. and Gaensler, B. M. and Michilli, D. and Masui, K. and Kaspi, V. M. and Su, J. and Bhardwaj, M. and Cassanelli, T. and Chawla, P. and Dong, F. (Adam) and Fonseca, E. and Leung, C. and Li, D. Z. and Ng, C. and Patel, C. and Pearlman, A. B. and Petroff, E. and Pleunis, Z. and Rafiei-Ravandi, M. and Rahman, M. and Sand, K. R. and Shin, K. and Stairs, I. H. and Tendulkar, S.},
title = {Revealing the Dynamic Magnetoionic Environments of Repeating Fast Radio Burst Sources through Multiyear Polarimetric Monitoring with CHIME/FRB},
journal = {The Astrophysical Journal}
}

@article{kumar2023,
    author = {Kumar, P and Luo, R and Price, D C and Shannon, R M and Deller, A T and Bhandari, S and Feng, Y and Flynn, C and Jiang, J C and Uttarkar, P A and Wang, S Q and Zhang, S B},
    title = {Spectropolarimetric variability in the repeating fast radio burst source FRB 20180301A},
    journal = {Monthly Notices of the Royal Astronomical Society},
    volume = {526},
    number = {3},
    pages = {3652-3672},
    year = {2023},
    month = {12},
    issn = {0035-8711},
    doi = {10.1093/mnras/stad2969},
    url = {https://doi.org/10.1093/mnras/stad2969},
    eprint = {https://academic.oup.com/mnras/article-pdf/526/3/3652/52050460/stad2969.pdf},
}

@article{abramson1982adaptive,
 ISSN = {00905364, 21688966},
 URL = {http://www.jstor.org/stable/2240724},
 author = {Ian S. Abramson},
 journal = {The Annals of Statistics},
 number = {4},
 pages = {1217--1223},
 publisher = {Institute of Mathematical Statistics},
 title = {On Bandwidth Variation in Kernel Estimates-A Square Root Law},
 urldate = {2026-09-09},
 volume = {10},
 year = {1982}
}

@article{Gordon_2025,
doi = {10.3847/1538-4357/ae0298},
url = {https://doi.org/10.3847/1538-4357/ae0298},
year = {2025},
month = {oct},
publisher = {The American Astronomical Society},
volume = {993},
number = {1},
pages = {119},
author = {Gordon, Alexa C. and Fong, Wen-fai and Deller, Adam T. and Marnoch, Lachlan and Lim, Sungsoon and Peng, Eric W. and Bannister, Keith W. and Bera, Apurba and Bhat, N. D. R. and Dial, Tyson and Dong, Yuxin and Eftekhari, Tarraneh and Glowacki, Marcin and Gourdji, Kelly and Gupta, Vivek and Jahns-Schindler, Joscha N. and Jaini, Akhil and Kilpatrick, Charles D. and Liu, Chang and Prochaska, J. Xavier and Ryder, Stuart D. and Shannon, Ryan M. and Simha, Sunil and Tejos, Nicolas and Wang, Yuanming and Wang, Ziteng},
title = {Mapping the Spatial Distribution of Fast Radio Bursts within their Host Galaxies},
journal = {The Astrophysical Journal}
}

@article{DELAIGLE2004249,
title = {Practical bandwidth selection in deconvolution kernel density estimation},
journal = {Computational Statistics & Data Analysis},
volume = {45},
number = {2},
pages = {249-267},
year = {2004},
issn = {0167-9473},
doi = {https://doi.org/10.1016/S0167-9473(02)00329-8},
url = {https://www.sciencedirect.com/science/article/pii/S0167947302003298},
author = {A. Delaigle and I. Gijbels}
}

@article{anderson1952asymptotic,
  title={Asymptotic theory of certain" goodness of fit" criteria based on stochastic processes},
  author={Anderson, Theodore W and Darling, Donald A},
  journal={The annals of mathematical statistics},
  pages={193--212},
  year={1952},
  publisher={JSTOR}
}

@article{sheather1990,
 ISSN = {01621459, 1537274X},
 URL = {http://www.jstor.org/stable/2289777},
 author = {Simon J. Sheather and J. S. Marron},
 journal = {Journal of the American Statistical Association},
 number = {410},
 pages = {410--416},
 publisher = {[American Statistical Association, Taylor & Francis, Ltd.]},
 title = {Kernel Quantile Estimators},
 urldate = {2026-09-12},
 volume = {85},
 year = {1990}
}

@ARTICLE{ravi2025,
       author = {{Ravi}, Vikram and {Catha}, Morgan and {Chen}, Ge and {Connor}, Liam and {Cordes}, James M. and {Faber}, Jakob T. and {Lamb}, James W. and {Hallinan}, Gregg and {Harnach}, Charlie and {Hellbourg}, Greg and {Hobbs}, Rick and {Hodge}, David and {Hodges}, Mark and {Law}, Casey and {Rasmussen}, Paul and {Sharma}, Kritti and {Sherman}, Myles B. and {Shi}, Jun and {Simard}, Dana and {Somalwar}, Jean J. and {Squillace}, Reynier and {Weinreb}, Sander and {Woody}, David P. and {Yadlapalli}, Nitika and {Deep Synoptic Array Team}},
        title = "{Deep Synoptic Array Science: A 50 Mpc Fast Radio Burst Constrains the Mass of the Milky Way Circumgalactic Medium}",
      journal = {\aj},
         year = 2025,
        month = jun,
       volume = {169},
       number = {6},
          eid = {330},
        pages = {330},
          doi = {10.3847/1538-3881/adc725},
archivePrefix = {arXiv},
       eprint = {2301.01000},
 primaryClass = {astro-ph.GA},
       adsurl = {https://ui.adsabs.harvard.edu/abs/2025AJ....169..330R}
}




\appendix

\section{Log-normal KDE and median calculation for RM}
\label{sec:KDE}


To calculate a KDE for the \rmeg\ distributions we first transform the \rmeg\ and its uncertainty into log$_{10}$ space:

\begin{equation}
\begin{split}
    & X_{\rm RM} = \log_{10}(|\rm RM_{EG}|) \\
    & \sigma_{X_{\rm RM}} = \frac{\sigma_{\rm RM}}{\rm RM_{EG}} \cdot\frac{1}{\ln(10)},
\end{split}
\end{equation}

where $X_{\rm RM}$ and $\sigma_{X_{\rm RM}}$ represent the absolute \rmeg\ and its uncertainty in log space respectively. The $X_{\rm RM}$ distribution is smoothed using a kernel density estimator (KDE) with a Gaussian kernel for each measurement $X_{\rm RM,i}$:

\begin{equation}
\label{eq:kernelgaussian}
    f_i(X_{\rm RM,i},h_i) = \sum_{i=1}^{n}\frac{1}{h_{i}\sqrt{2\pi}}\exp\Bigg(\frac{-\big(X_{\rm RM}-X_{\rm RM,i}\big)^{2}}{2h_{i}^{2}}\Bigg),
\end{equation}

where n is the number of samples in the discrete distribution and $h_{i}$ is a smoothness-aware KDE bandwidth used for kernel smoothing. The full normalised KDE is then:

\begin{equation}
\begin{split}
    &f(X_{\rm RM}) =  \sum_{i=1}^{n}f_{i}(X_{\rm RM,i}),\\
    &{f}(X_{\rm RM}) \rightarrow \frac{{f}(X_{\rm RM})}{{\int_{-\infty}^{\infty}{\big[{f}(X_{\rm RM})}}\big]d\log_{10}(\rm |RM_{EG}|)}.
    \end{split}
\end{equation}

The bandwidth $h_{i}$ is calculated by first estimating the intrinsic variance in $X_{\rm RM}$ \citep{DELAIGLE2004249}:

\begin{equation}
    \hat{\sigma}^{2}_{X_{\rm RM}} = \frac{\sum_{i=1}^{n}{(X_{\rm RM} - \bar{X}_{\rm RM})^2}}{n-1} - \frac{1}{n}\sum{\sigma_{X_{\rm RM,i}}}^2,
\end{equation}

where the first term represents on the right hand side of the equation represents the measured variance in $X_{\rm RM}$. Using Silverman's rule \citep{silverman2018density} a base bandwidth $h$ is calculated:

\begin{equation}
\label{eq:base_h}
    h = 1.06\cdot\hat{\sigma}_{X_{\rm RM}}n^{\alpha},
\end{equation}

where $\alpha$ is the bandwidth scaling exponent. For smoothing the $X_{\rm RM}$ distribution an exponent of $\alpha$ = -1/5 is used, which minimises the mean integrated squared error. The measurement uncertainties are then systematically added back into the base bandwidth to calculate a pilot bandwidth:

\begin{equation}
\label{eq:pilot_h}
    h(X_{\rm RM,i}) = \sqrt{h^2 + \sigma^{2}_{X_{\rm RM},i}}.
\end{equation}

This pilot bandwidth is used to estimate an adaptive scale factor $\lambda_i$ which ensures appropriate smoothing in the $X_{\rm RM}$ distribution, which has regions of high density and low density. Using Abramson's law \citep{abramson1982adaptive}, $\lambda_i$ is estimated as:

\begin{equation}
    \lambda_{i} = \Bigg(\frac{f_i(X_{\rm RM,i}, h(X_{\rm RM,i}))}{\exp\big(\sum_{i=1}^n{\ln(f_i(X_{\rm RM,i},h(X_{\rm RM,i})}\big)}\Bigg)^{-1/2},
\end{equation}

where $f_i(X_{\rm RM,i},h(X_{\rm RM,i}))$ is calculated from equation~\ref{eq:kernelgaussian} using the pilot bandwidth from equation~\ref{eq:pilot_h} for each measurement $X_{\rm RM,i}$. Finally, the adaptive bandwidth is calculated using the following heuristic:

\begin{equation}
    h_i = \max(\lambda_ih, 0.6\cdot h),
\end{equation}

where $h$ is the base bandwidth calculated using equation~\ref{eq:base_h} and 0.6$h$ is used as the minimum bandwidth in smoothing the $X_{\rm RM}$ distribution. The log-normal KDE for each FRB sample are plotted in Fig.~\ref{fig:rmhist} (panel A).

Both the arithmetic median and geometric mean along with their uncertainties are estimated for each sample. The uncertainty in the median is estimated using a bootstrapping routine. First, the distribution of $X_{\rm RM}$ is sampled with replacement along with the adaptive bandwidth $h_{i}$ using a uniform probability distribution to get a new distribution $Y_{\rm RM}$:

\begin{equation}
\label{eq:bootstrap_sample}
    Y_{\rm RM,i} = X^{*}_{\rm RM,i} + \mu,\hspace{0.1cm}\mathrm{where}\hspace{0.1cm} \mu \sim \mathcal{N}\big(0,(h^{*}_{m,i})^2\big),
\end{equation}

where the `*' notation denotes a sample drawn from the original distribution with replacement using a uniform probability distribution. $\mu$ is a noise term drawn from a normal probability distribution with a standard deviation equal to the adaptive bandwidth $h_{m,i}$. It is important to note that this adaptive bandwidth is slightly different to the bandwidth used in the KDE as noted by the subscript $m$. The optimal bandwidth for estimating the median scales as $h\propto n^{-1/3}$ \citep{sheather1990}. Hence, $h_{m}$ is calculated in almost the same way as the adaptive KDE bandwidth, just with an exponent of $\alpha$ = -1/3 in equation~\ref{eq:base_h}.

Constructing the bootstrap sample $Y_{\rm RM,i}$ using equation~\ref{eq:bootstrap_sample} inflates the total variance. To mitigate this, a correction scalar is applied to sample:

\begin{equation}
    Y_{\rm rescaled} = \bar{X}_{\rm RM}^{*} + \frac{Y_{\rm RM} - \bar{X}_{\rm RM}^{*}}{\sqrt{1 + \frac{\text{Var}(h^{*}_{m})}{\text{Var}(\sigma_{X_{\rm RM}}^{*})}}},
\end{equation}

where $\text{Var}$ denotes the variance and $\bar{X}_{\rm RM}^{*}$ is the mean of the bootstrapped sample. The arithmetic median for the bootstrapped sample is given by:

\begin{equation}
    M_{s} = \mathrm{med}\big(10^{Y_{\rm rescaled}}\big).
\end{equation}

The geometric mean of the bootstrapped sample is straightforward to calculate:

\begin{equation}
    G_{s} = 10^{\frac{1}{n}\sum_{i=1}^{n}{X^{*}_{\rm RM, i}}}.
\end{equation}

1000 bootstraps are performed on the $X_{\rm RM}$ distribution. The 68\% confidence intervals are estimated from the set of bootstrapped samples for the median and geometric mean. The median and geometric mean of the data is calculated as follows:

\begin{equation}
\begin{split}
    &M_{\rm |RM_{EG}|} = \rm med(|RM_{EG}|)\\
    &G_{\rm |RM_{EG}|} = \bigg(\prod_{i=1}^{n}\rm|RM_{EG,i}|\bigg)^{\frac{1}{n}}.
\end{split}
\end{equation}

The upper and lower bounds on both the median and geometric mean are calculated are from the 84th and 16th percentiles of the bootstrapped sets $M_{s}$ and $G_{s}$ respectively. These values are reported in Table~\ref{tab:RMtable}.

As a consistency check, we calculated the geometric mean and uncertainty directly from the $\log_{10}(\rm|RM|)$ data. This yields a geometric mean of 50.02$^{+9.17}_{-7.75}$ rad m$^{-2}$ and 91.02$^{+23.34}_{-17.48}$ rad m$^{-2}$ for the CHIME and ASKAP distributions respectively; both the mean and uncertainty closely match the more involved method detailed above.

\section{ASKAP Data}

\renewcommand{\thetable}{B\arabic{table}}
\setcounter{table}{0}

Table~\ref{tab:burstprop} reports the position, redshift and burst properties of the \newhtr{} sample. Table~\ref{tab:burstpol} summarises the polarisation properties of the \newhtr{} bursts. Finally, some  updated burst properties of the \oldhtr{} sample are reported in Table~\ref{tab:htr1_prop}. 

\begin{table*}
    \centering
    \setlength\extrarowheight{3pt}
    \setlength\tabcolsep{3pt}
    \begin{threeparttable}
    \caption{Burst properties of \newhtr{}. S/N is the integrated S/N of the burst over its width $w_{95}$ and bandwidth $\Delta\nu_{95}$. DM is the structure-maximised DM. We apply a flat and conservative uncertainty of 20\% for both fluence and spectral luminosity.}
    \label{tab:burstprop}
    \begin{tabular}{ccccccccccccc}
    \hline
FRB & \thead{RA\\(J2000)} & \thead{DEC\\(J2000)} & \thead{$\rm \sigma_{RA}$\\(\arcs)} & \thead{$\rm \sigma_{DEC}$\\(\arcs)} & $z$ & \thead{$w_{95}$\\(ms)} & S/N & \thead{DM\\(pc cm$^{-3}$)} & \thead{$\nu_{c}$\\(MHz)} & \thead{$\Delta\nu_{95}$\\(MHz)} & \thead{fluence\\(Jy ms)} & \thead{spectral luminosity\\(10$^{34}$ ergs s$^{-1}$ Hz$^{-1}$)} \\ \hline
        20240525A & 00:27:07.89 & -06:53:22.65 & 0.52 & 0.47 & 0.327\tnote{c} & 21.68 & 105.1 & 491.23$^{+1.19}_{-0.82}$ & 943 & 244 & 108.8$\pm$21.8 & 1.43$\pm$0.29 \\ 
        20240629F & 05:34:11.54 & -59:40:25.16 & 0.70 & 0.55 &  & 5.42 & 35.1 & 534.78$^{+0.26}_{-0.28}$ & 832 & 310 & 17.7$\pm$3.5 &  \\ 
        20240807A & 01:33:12.77 & -06:59:03.20 & 0.87 & 0.58 & 0.2098\tnote{d} & 30.8 & 16.2 & 518.10$^{+3.50}_{-3.60}$ & 973 & 226 & 20.8$\pm$4.2 & 0.08$\pm$0.02 \\ 
        20241014B & 03:19:21.38 & -58:19:15.70 & 0.62 & 0.58 & 0.3355\tnote{d} & 1.6 & 71.3 & 497.66$^{+0.22}_{-0.90}$ & 896 & 272 & 18.3$\pm$3.7 & 3.43$\pm$0.69 \\ 
        20241027B & 02:24:06.92 & -20:42:12.21 & 0.62 & 0.49 & 0.336\tnote{c} & 43.6 & 44.9 & 323.00$^{+14.50}_{-12.70}$ & 933 & 134 & 84.3$\pm$16.9 & 0.58$\pm$0.12 \\ 
        20241226C & 04:00:34.64 & -55:33:20.71 & 0.68 & 0.49 &  & 19.2 & 15.2 & 647.40$^{+3.00}_{-2.80}$ & 933 & 278 & 13.0$\pm$2.6 &  \\ 
        20250106B & 03:37:10.97 & -16:12:16.17 & 0.83 & 0.56 & 0.3475\tnote{e} & 8.56 & 50.6 & 578.01$^{+0.30}_{-0.31}$ & 849 & 336 & 33.3$\pm$6.7 & 1.25$\pm$0.25 \\ 
        20250201A & 14:13:34.21 & -74:32:59.04 & 0.59 & 0.52 &  & 13.5 & 26.5 & 961.80$^{+2.10}_{-2.00}$ & 991 & 193 & 19.2$\pm$3.8 &  \\ 
        20250313A & 04:22:53.77 & -52:16:11.23 & 0.55 & 0.51 &  & 2.65 & 28.2 & 791.30$^{+2.90}_{-2.90}$ & 927 & 317 & 8.0$\pm$1.6 &  \\ 
        20250511A & 01:09:32.79 & -44:31:37.89 & 0.66 & 0.53 & 0.218\tnote{e} & 0.67 & 48.0 & 360.72$^{+0.06}_{-0.06}$ & 930 & 308 & 9.7$\pm$1.9 & 1.77$\pm$0.35 \\ 
        20250520A & 11:50:53.88 & -45:12:49.15 & 0.44 & 0.44 & 0.1685\tnote{e} & 4.46 & 42.5 & 341.66$^{+0.30}_{-0.30}$ & 1278 & 320 & 15.0$\pm$3.0 & 0.24$\pm$0.05 \\ 
        20250607C & 02:32:43.45 & -39:20:27.94 & 0.42 & 0.41 & 0.2106\tnote{e} & 9.04 & 264.5 & 335.87$^{+0.34}_{-0.26}$ & 924 & 315 & 174.4$\pm$34.9 & 2.21$\pm$0.44 \\ 
        20250613A\tnote{a} & 04:36:33.04 & -44:31:58.57 & 0.55 & 0.44 &  &  &  &  &  &  &  &  \\ 
        20251010B & 15:35:58.02 & -65:48:15.92 & 0.61 & 0.61 &  & 680 & 20.0 & 1033.50$^{+0.20}_{-0.20}$ & 843 & 305 & 111.5$\pm$22.3 &  \\ 
        20251019B & 10:08:27.45 & +16:23:56.36 & 0.54 & 0.56 &  & 5.35 & 26.5 & 1277.15$^{+0.06}_{-0.06}$ & 875 & 311 & 10.9$\pm$2.2 &  \\ 
        20251024B & 02:29:30.46 & -62:57:46.24 & 0.49 & 0.44 & 0.319\tnote{e} & 12.45 & 50.4 & 543.54$^{+0.21}_{-0.21}$ & 882 & 299 & 34.2$\pm$6.8 & 0.74$\pm$0.15 \\ 
        20251026B\tnote{a} & 04:36:32.98 & -44:31:58.79 & 0.43 & 0.42 &  &  &  &  &  &  &  &  \\ 
        20251119B & 04:50:24.11 & -59:59:25.92 & 0.46 & 0.45 &  & 1.095 & 162.5 & 735.84$^{+0.02}_{-0.03}$ & 912 & 336 & 31.4$\pm$6.3 &  \\ 
        20251214A & 20:22:13.21 & -80:10:49.21 & 0.77 & 0.91 &  & 33 & 14.4 & 327.74$^{+0.74}_{-0.74}$ & 815 & 268 & 19.1$\pm$3.8 &  \\ 
        20260110C & 01:17:32.74 & +18:01:18.32 & 0.46 & 0.46 &  & 4.7 & 54.7 & 488.83$^{+1.50}_{-2.30}$ & 844 & 270 & 26.5$\pm$5.3 &  \\ 
        20260213T & 11:03:32.97 & +04:15:12.32 & 0.42 & 0.43 &  & 3.76 & 57.6 & 861.13$^{+0.37}_{-0.27}$ & 931 & 314 & 17.4$\pm$3.5 &  \\ 
        20260217D\tnote{b} & 19:12:41.20 & -65:34:48.28 & 0.63 & 0.51 &  & 3.5 & 30.4 & 1177.65$^{+0.85}_{-0.85}$ & 934 & 308 & 10.0$\pm$2.0 &  \\ 
        20260307D\tnote{b} & 19:51:03.57 & -64:59:47.58 & 1.40 & 1.47 &  & 6.36 & 30.6 & 1551.22$^{+0.72}_{-0.72}$ & 919 & 318 & 22.4$\pm$4.5 &  \\ 
        20260324E & 04:15:34.38 & +08:26:35.55 & 0.49 & 0.47 & 0.2475 & 85 & 36.6 & 343.30$^{+4.80}_{-4.80}$ & 865 & 268 & 72.8$\pm$14.6 & 0.14$\pm$0.03 \\ 
        20260326D & 21:18:54.37 & -19:16:06.17 & 0.52 & 0.47 &  & 14.44 & 21.3 & 494.74$^{+0.01}_{0.01}$ & 811 & 248 & 18.9$\pm$3.8 &  \\ 
        20260326E & 21:16:06.32 & -17:57:26.56 & 0.56 & 0.48 &  & 16.8 & 33.0 & 582.07$^{+1.80}_{-1.80}$ & 826 & 302 & 29.0$\pm$5.8 &  \\ 
        20260331E & 23:06:48.64 & -58:12:04.59 & 0.47 & 0.44 & 0.21 & 4.5 & 49.5 & 307.99$^{+1.20}_{-1.70}$ & 790 & 253 & 28.7$\pm$5.7 & 0.73$\pm$0.15 \\ 
        20260409E & 23:16:51.40 & -57:37:03.04 & 0.65 & 0.55 &  & 250 & 14.4 & 657.90\tnote{f} & 887 & 262 & 42.8$\pm$8.6 &  \\ 
        20260417A & 23:37:53.23 & -49:49:17.33 & 0.59 & 0.57 &  & 20 & 29.2 & 724.53$^{+1.83}_{-2.48}$ & 823 & 121 & 33.4$\pm$6.7 &  \\ 
        20260417B & 14:58:54.83 & +00:09:46.05 & 0.45 & 0.45 &  & 5.9 & 25.9 & 382.04$^{+0.04}_{-0.04}$ & 906 & 303 & 11.2$\pm$2.2 &  \\ 
        20260515J & 14:27:21.94 & -19:24:54.78 & 0.49 & 0.45 &  & 18.8 & 36.1 & 292.12$^{+1.70}_{-1.70}$ & 1272 & 336 & 25.6$\pm$5.1 &  \\ 
        20260608D & 10:14:27.14 & +18:18:33.36 & 0.42 & 0.43 & 0.113 & 1.92 & 234.7 & 331.05$^{+0.10}_{-0.10}$ & 864 & 336 & 61.3$\pm$12.3 & 1.02$\pm$0.20 \\ 
        20260614A & 16:57:43.51 & -52:11:03.70 & 0.72 & 0.57 &  & 15.2 & 33.9 & 690.34$^{+4.00}_{-5.00}$ & 937 & 301 & 25.7$\pm$5.1 &  \\ 
        20260615E & 13:25:54.86 & -10:40:00.63 & 0.53 & 0.47 &  & 2.28 & 49.0 & 791.80$^{+0.09}_{-0.08}$ & 907 & 308 & 12.9$\pm$2.6 &  \\ \hline
    \end{tabular}
    \begin{tablenotes}
      \small
      \item[a] Bursts from repeating FRB 20250613A, see \cite{dial2026} for burst properties.
      \item[b] Only recovered a single polarisation beam.
      \item[c] \cite{Gordon_2025}
      \item[d] Y. Wang et al. in prep
      \item[e] A. C. Gordon et al. in prep
      \item[f] Detection DM is reported since an optimal DM could not be measured using structure-maximisation \citep{sutinjo2023calculation} due to the low S/N and large width.
    \end{tablenotes}
    \end{threeparttable}
\end{table*}

\begin{table*}
    \centering
    \setlength\extrarowheight{3pt}
    \setlength\tabcolsep{12pt}
    \begin{threeparttable}
    \caption{Polarisation properties of \newhtr. $\nu_{0}$ is the power-averaged central frequency of the burst measured using \texttt{RM-tools}.}
    \label{tab:burstpol}
    \begin{tabular}{ccccccc}
    \hline
        FRB & \thead{RM\\(rad m$^{-2}$)} & \thead{$\nu_{0}$\\(MHz)} & L/I & |V|/I & P/I & \thead{RM$_{MW}$\\(rad m$^{-2}$)} \\ \hline
        20240525A & -252.3$\pm$1.9 & 947 & 0.17$\pm$0.01 & 0.03$\pm$0.01 & 0.17$\pm$0.01 & 18.2$\pm$0.9 \\ 
        20240629F & 36.4$\pm$0.5 & 814 & 1.01$\pm$0.04 & 0.12$\pm$0.03 & 1.02$\pm$0.04 & 43.4$\pm$1.5 \\ 
        20240807A &  &  & $\leq$0.37 &  &  & 1.2$\pm$2.6 \\ 
        20241014B & 59.0$\pm$0.4 & 918 & 0.96$\pm$0.02 & 0.03$\pm$0.01 & 0.96$\pm$0.02 & 47.8$\pm$1.2 \\ 
        20241027B &  & 953 & $\leq$0.14 & 0.65$\pm$0.03 & $\leq$0.65 & 15.2$\pm$1.4 \\ 
        20241226C &  &  & $\leq$0.39 &  &  & 21.4$\pm$1.1 \\ 
        20250106B & -403.4$\pm$0.5 & 874 & 0.76$\pm$0.02 & 0.16$\pm$0.02 & 0.76$\pm$0.02 & 58.5$\pm$3.1 \\ 
        20250201A & -86.9$\pm$2.0 & 1032 & 0.93$\pm$0.05 & 0.13$\pm$0.04 & 0.94$\pm$0.05 & -56.9$\pm$6.6 \\ 
        20250313A &  &  & $\leq$0.21 &  &  & 2.5$\pm$1.4 \\ 
        20250511A & -365.1$\pm$0.5 & 1020 & 0.84$\pm$0.03 & 0.23$\pm$0.02 & 0.87$\pm$0.03 & -2.0$\pm$2.5 \\ 
        20250520A & -134.9$\pm$2.0 & 1255 & 0.84$\pm$0.03 & 0.11$\pm$0.02 & 0.85$\pm$0.03 & -81.2$\pm$4.2 \\ 
        20250607C & -427.0$\pm$0.1 & 918 & 0.74$\pm$0.01 & 0.11$\pm$0.01 & 0.75$\pm$0.01 & 6.9$\pm$2.3 \\ 
        20250613A\tnote{a} &  &  &  &  &  & 12.4$\pm$1.2 \\ 
        20251010B &  &  & $\leq$0.30 &  &  & 41.1$\pm$7.4 \\ 
        20251019B & -20.2$\pm$1.2 & 875 & 0.74$\pm$0.05 & 0.18$\pm$0.04 & 0.77$\pm$0.05 & 17.0$\pm$1.0 \\ 
        20251024B & 1022.8$\pm$0.9 & 911 & 0.44$\pm$0.02 & 0.05$\pm$0.02 & 0.45$\pm$0.02 & 20.1$\pm$1.2 \\ 
        20251026B\tnote{a} &  &  &  &  &  & 12.4$\pm$1.2 \\ 
        20251119B & 77.5$\pm$0.1 & 889 & 0.97$\pm$0.01 & 0.03$\pm$0.01 & 0.97$\pm$0.01 & 58.3$\pm$3.3 \\ 
        20251214A &  &  & $\leq$0.42 &  &  & 25.7$\pm$2.1 \\ 
        20260110C & -644.6$\pm$0.5 & 882 & 0.76$\pm$0.02 & 0.07$\pm$0.02 & 0.76$\pm$0.02 & -30.1$\pm$1.2 \\ 
        20260213T & 258.6$\pm$0.5 & 916 & 0.78$\pm$0.02 & 0.26$\pm$0.02 & 0.82$\pm$0.02 & 18.8$\pm$2.0 \\ 
        20260217D\tnote{b} &  &  &  &  &  & 36.9$\pm$1.1 \\ 
        20260307D\tnote{b} &  &  &  &  &  & 14.3$\pm$1.7 \\ 
        20260324E & -926.9$\pm$2.3 & 880 & 0.35$\pm$0.03 & 0.11$\pm$0.04 & 0.37$\pm$0.03 & 5.4$\pm$3.8 \\ 
        20260326D & -5.8$\pm$0.6 & 807 & 1.00$\pm$0.04 & 0.10$\pm$0.03 & 1.00$\pm$0.04 & -3.0$\pm$1.4 \\ 
        20260326E & -118.0$\pm$0.6 & 827 & 0.92$\pm$0.04 & 0.07$\pm$0.03 & 0.93$\pm$0.04 & -3.4$\pm$1.8 \\ 
        20260331E & 48.1$\pm$1.8 & 719 & 0.25$\pm$0.02 & 0.26$\pm$0.02 & 0.36$\pm$0.02 & 5.9$\pm$2.0 \\ 
        20260409E &  &  & $\leq$0.42 &  &  & 19.3$\pm$1.2 \\ 
        20260417A & 225.6$\pm$2.2 & 821 & 1.08$\pm$0.07 & 0.17$\pm$0.05 & 1.09$\pm$0.07 & 5.8$\pm$1.9 \\ 
        20260417B & -19.9$\pm$1.0 & 902 & 0.88$\pm$0.05 & 0.13$\pm$0.04 & 0.89$\pm$0.05 & 4.7$\pm$1.1 \\ 
        20260515J &  &  & $\leq$0.17 &  &  & -7.9$\pm$1.3 \\ 
        20260608D & -125.0$\pm$0.2 & 889 & 0.48$\pm$0.01 & 0.55$\pm$0.01 & 0.20$\pm$0.01 & 0.0$\pm$1.2 \\ 
        20260614A & -279.6$\pm$4.1 & 957 & 0.36$\pm$0.03 & 0.08$\pm$0.03 & 0.37$\pm$0.03 & 28.7$\pm$25.4 \\ 
        20260615E & 139.7$\pm$1.0 & 884 & 0.62$\pm$0.03 & 0.16$\pm$0.03 & 0.64$\pm$0.03 & 7.0$\pm$0.8 \\ \hline
    \end{tabular}
    \begin{tablenotes}
      \small
      \item[a] Bursts from repeating FRB 20250613A; see \cite{dial2026} for polarisation properties.
      \item[b] Only a single polarisation was saved for this burst due to a malfunction of the voltage download, meaning no polarisation properties could be recovered.
    \end{tablenotes}
    \end{threeparttable}
\end{table*}  

\begin{table*}
    \centering
    \setlength\extrarowheight{3pt}
    \setlength\tabcolsep{5pt}
    \begin{threeparttable}
    \caption{Updated properties of \oldhtr{} published in \protect\cite{scott2025high}. $\nu_{c}$ is the central frequency and $\Delta\nu_{95}$ is the bandwidth that incorporates 95\% of the burst fluence as outlined in Section~\ref{sec:widthfitting}. Bursts with no spectroscopically confirmed host have no luminosity entry, and FRB 20190102C has no estimated fluence due to an uncertain primary beam correction.}
    \label{tab:htr1_prop}
    \begin{tabular}{cccccc}
    \hline
        FRB  & \thead{$\nu_{c}$\\(MHz)} & \thead{$\Delta\nu_{95}$\\(MHz)} & \thead{RM$_{MW}$\\(rad m$^{-2}$)} & \thead{fluence\\(Jy ms)} & \thead{spectral luminosity\\(10$^{34}$ ergs s$^{-1}$ Hz$^{-1}$)} \\ \hline
        20180924B & 335 & 1319.5 & 22.4$\pm$0.9 & 14.7$\pm$2.9 & 2.02$\pm$0.40 \\ 
        20181112A\tnote{a} & 336 & 1297.5 & 30.1$\pm$1.1 & 29.2$\pm$5.8 & 22.50$\pm$4.50 \\ 
        20190102C & 336 & 1271.5 & 32.4$\pm$1.5 &  &  \\ 
        20190608B & 335 & 1271 & -22.1$\pm$2.5 & 16.6$\pm$3.3 & 0.05$\pm$0.01 \\ 
        20190611B & 271 & 1239 & 29.1$\pm$1.3 & 10.0$\pm$2.0 & 2.40$\pm$0.48 \\ 
        20190711A & 336 & 1271.5 & 11.6$\pm$1.5 & 17.9$\pm$3.6 & 1.22$\pm$0.24 \\ 
        20190714A & 336 & 1271.5 & -11.4$\pm$2.4 & 12.2$\pm$2.4 & 0.59$\pm$0.12 \\ 
        20191001A & 336 & 919.5 & 22.2$\pm$1.1 & 66.1$\pm$13.2 & 0.67$\pm$0.13 \\ 
        20191228B & 336 & 1271.5 & 6.8$\pm$2.5 & 65.7$\pm$13.1 & 0.74$\pm$0.15 \\ 
        20200430A & 336 & 863.5 & 25.4$\pm$1.3 & 62.9$\pm$12.6 & 0.18$\pm$0.04 \\ 
        20200906A & 296 & 843.5 & 81.1$\pm$10.6 & 90.3$\pm$18.1 & 257.24$\pm$51.45 \\ 
        20210117A & 336 & 1271.5 & 2.0$\pm$1.8 & 29.2$\pm$5.8 & 0.96$\pm$0.19 \\ 
        20210320C & 264 & 827.5 & -5.0$\pm$1.1 & 70.2$\pm$14.0 & 16.37$\pm$3.27 \\ 
        20210407E & 228 & 1217.5 & -33.4$\pm$5.7 & 53.8$\pm$10.8 &  \\ 
        20210912A & 316 & 1261.5 & -2.1$\pm$0.9 & 65.8$\pm$13.2 &  \\ 
        20211127I & 336 & 1271.5 & 5.0$\pm$1.2 & 30.9$\pm$6.2 & 0.34$\pm$0.07 \\ 
        20211203C & 336 & 919.5 & -36.1$\pm$2.1 & 45.5$\pm$9.1 & 0.56$\pm$0.11 \\ 
        20211212A & 336 & 1631.5 & 7.6$\pm$2.9 & 81.1$\pm$16.2 & 0.18$\pm$0.04 \\ 
        20220105A & 336 & 1631.5 & -1.5$\pm$0.8 & 14.2$\pm$2.8 & 1.29$\pm$0.26 \\ 
        20220501C & 336 & 863.5 & 14.4$\pm$0.9 & 52.4$\pm$10.5 & 2.96$\pm$0.59 \\ 
        20220610A & 75 & 1141 & 11.0$\pm$1.0 & 55.1$\pm$11.0 & 78.53$\pm$15.71 \\ 
        20220725A & 336 & 919.5 & 7.4$\pm$0.9 & 68.2$\pm$13.6 & 0.81$\pm$0.16 \\ 
        20220918A & 60 & 1133.5 & 35.9$\pm$2.8 & 106.0$\pm$21.2 & 5.04$\pm$1.01 \\ 
        20221106A & 336 & 1631.5 & 49.4$\pm$1.0 & 98.1$\pm$19.6 & 1.53$\pm$0.31 \\ 
        20230526A & 336 & 1271.5 & 5.1$\pm$2.1 & 35.2$\pm$7.0 & 0.81$\pm$0.16 \\ 
        20230708A & 336 & 919.5 & 42.9$\pm$2.4 & 107.9$\pm$21.6 & 0.13$\pm$0.03 \\ 
        20230718A & 336 & 1271.5 & 265.2$\pm$12.4 & 14.0$\pm$2.8 & 0.06$\pm$0.01 \\ 
        20230731A & 336 & 1271.5 & 86.2$\pm$11.2 & 14.5$\pm$2.9 &  \\ 
        20230902A & 300 & 849.5 & 13.4$\pm$1.1 & 28.3$\pm$5.7 & 14.61$\pm$2.92 \\ 
        20231226A & 336 & 863.5 & 8.7$\pm$1.6 & 114.6$\pm$22.9 &  \\ 
        20240201A & 331 & 917 & 4.3$\pm$1.1 & 43.6$\pm$8.7 & 0.05$\pm$0.01 \\ 
        20240208A & 336 & 863.5 & -1.4$\pm$1.5 & 21.6$\pm$4.3 &  \\ 
        20240210A & 336 & 863.5 & -13.7$\pm$1.1 & 131.9$\pm$26.4 & 0.12$\pm$0.02 \\ 
        20240304A & 256 & 879.5 & -53.3$\pm$2.4 & 53.8$\pm$10.8 &  \\ 
        20240310A & 178 & 840.5 & 4.6$\pm$2.4 & 34.6$\pm$6.9 & 0.10$\pm$0.02 \\ 
        20240318A & 336 & 919.5 & 14.5$\pm$1.1 & 15.9$\pm$3.2 &  \\ \hline
    \end{tabular}
    \begin{tablenotes}
      \small
      \item[a] \cite{cho2020spectropolarimetric}.
    \end{tablenotes}
    \end{threeparttable}
\end{table*}

\section{$\rm |RM_{EG}$| as a function of L/I}

Figure.~\ref{fig:rm_v_li} shows a scatter plot of extragalactic |RM| as a function of the linear polarisation fraction for both the CHIME and ASKAP burst catalogues with measurable RMs.

\renewcommand{\thefigure}{C\arabic{figure}}
\setcounter{figure}{0}

\begin{figure}
    \centering
    \includegraphics[width=1.0\linewidth]{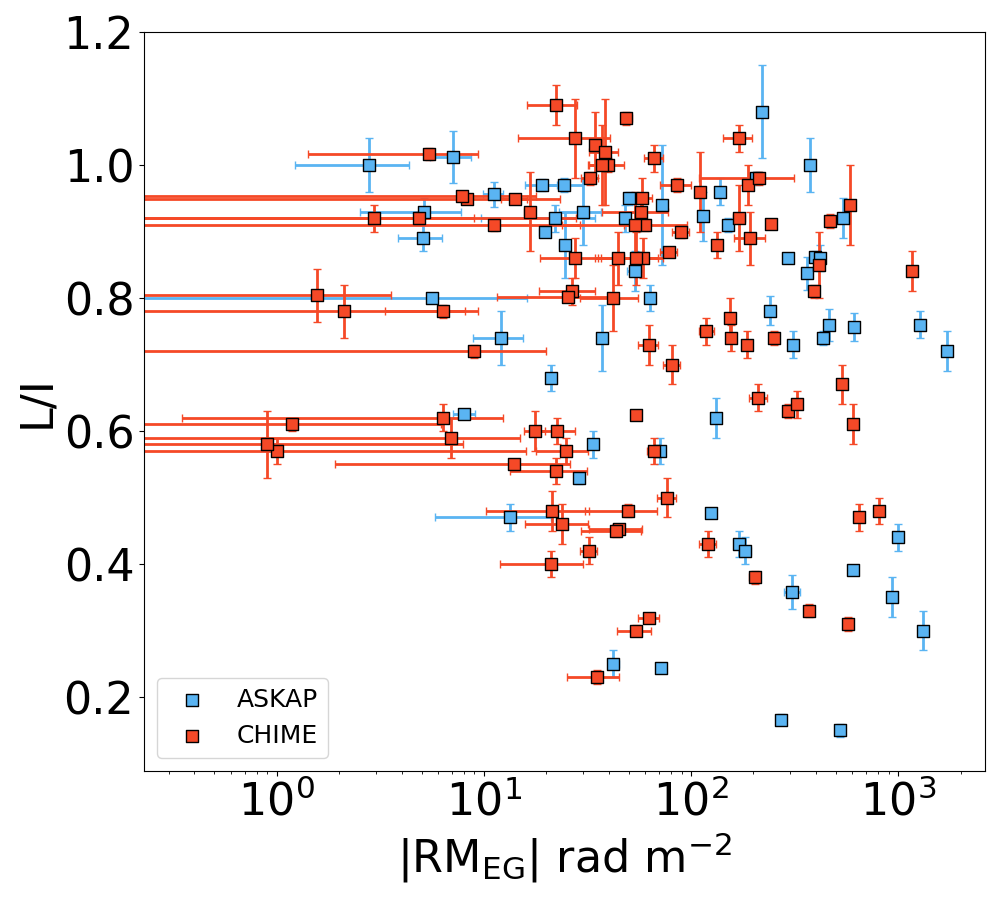}
    \caption{Extragalactic RM as a function of L/I. Spearman rank coefficient for CHIME sample is -0.052 \citep{pandhi2024polarization} and the coefficient for the ASKAP sample is -0.345. }
    \label{fig:rm_v_li}
\end{figure}


\bsp	
\label{lastpage}
\end{document}